\documentclass{article}
\usepackage{amsmath,amsfonts}
\usepackage{graphicx}
\usepackage{grffile}
\input epsf
\usepackage[nosort]{cite}
\usepackage{amssymb,amsthm}
\usepackage{breqn}
\usepackage{enumerate}
\usepackage{tensor}
\usepackage{mathrsfs}
\usepackage{comment}

\usepackage{xcolor}

\usepackage[english]{babel}
\usepackage{braket}
\usepackage{dcolumn}
\usepackage{bm}
\usepackage[hidelinks]{hyperref}

\makeatletter\@addtoreset{equation}{section}\makeatother

\def\bra#1{\mathinner{\langle{#1}|}}
\def\ket#1{\mathinner{|{#1}\rangle}}

\renewcommand{\title}[1]{\vbox{\center\LARGE{#1}}\vspace{5mm}}
\renewcommand{\author}[1]{\vbox{\center#1}\vspace{5mm}}
\newcommand{\address}[1]{\vbox{\center\em#1}}

\newcommand{\tr}{{\rm{tr~}}}
\usepackage{tocloft}

\begin{document}

\newtheorem{theorem}{Theorem}
\newtheorem{corollary}{Corollary}
\newtheorem{conjecture}{Conjecture}
\begin{titlepage}
\begin{center}
\vskip 1cm

\title{A Search for High-Threshold Qutrit Magic State Distillation Routines}

\author{Shiroman Prakash and Rishabh Singhal}

\address{Department of Physics and Computer Science, Dayalbagh Educational Institute, Agra, India}

\end{center}

\begin{abstract}
Determining the best attainable threshold for qudit magic state distillation is directly related to the question of whether or not contextuality is sufficient for universal quantum computation. We show that the performance of a qudit correcting code for magic state distillation is captured by its complete weight enumerator. For the qutrit “strange” 
 state -- a maximally magic non-stabilizer state -- the performance of a code is captured by its simple weight enumerator. This result allows us to carry out an extensive search for high-threshold magic state distillation routines for the strange state. Our search covers all $[[n,1]]_3$ qutrit stabilizer codes with a complete set of transversal Clifford gates for $n\leq 23$, and all $[[n,1]]_3$ stabilizer codes with a transversal $H^2$ gate with $n \leq 9$ qudits. For $n=23$, we find over 600 CSS codes that can distill the qutrit strange state with cubic noise suppression. While none of these codes surpass the threshold of the 11-qutrit Golay code, their existence suggests that, for large codes, the ability to distill the qutrit strange state is somewhat generic. 
\end{abstract}

\vfill

\end{titlepage}

\eject \tableofcontents

\clearpage
\section{Introduction}

Contextuality was identified as a necessary and possibly sufficient condition for universal quantum computing in \cite{nature}. The argument of \cite{nature} is based on magic state distillation \cite{MSD} for qudits of odd-prime dimension, and later extended to qudits of arbitrary odd dimensions in \cite{Delfosse_2017}, and continuous variable systems in \cite{haferkamp2021equivalence}.\footnote{There are certain subtleties associated with state-independent contextuality for qubits and qudits of even dimension -- for simplicity, we focus exclusively on qudits of odd-prime dimension in this paper.} The authors of these works showed that qudit states that do not exhibit contextuality with respect to stabilizer measurements have a non-negative discrete Wigner function \cite{Gross, Wootters1987}. The set of such states is known as the Wigner polytope. Because Clifford unitaries and stabilizer measurements are efficiently simulable for states in the Wigner polytope \cite{Veitch_2012, Veitch_2014}, they thus cannot be distilled into pure magic states. While this argument shows that contextuality is necessary for universal quantum computation, the conjecture that contextuality is sufficient for quantum computation remains open.

In order to demonstrate that contextuality is not only necessary but also sufficient for qutrit quantum computation, one must demonstrate that a supply of qudits that do exhibit contextuality with respect to stabilizer measurements may be used to achieve universal quantum computation. In the language of magic state distillation, this translates into the question, can any qudit mixed state outside the Wigner polytope be distilled into a pure magic state? 

For qudits of odd prime dimension $p$, the Wigner polytope is a convex polytope with $p^2$ facets that lives in the $p^2-1$ dimensional space of qudit density matrices.  In \cite{prakash2020contextual}, it was shown that no finite magic state distillation routine can distill all states that lie outside one of the faces of the Wigner polytope\footnote{Some distillation routines that distill qutrit states up to one of the hyperedges of the Wigner polytope were found in \cite{DawkinsHoward}.}, generalizing the analogous result for qubit states that lie outside the stabilizer polytope \cite{campbell2010bound}. However, the possibility remains that a sequence of magic state distillation routines, based on stabilizer codes of increasing length $n$ may distill states arbitrarily close to a face of the Wigner polytope. Is there any evidence that such a sequence of magic state distillation routines exists?

The problem simplifies if one focuses on qutrits. There exists a qutrit magic state, first identified by Howard and van Dam \cite{HowardVanDam}, sometimes known in the literature in the qutrit strange state $\ket{S}$ \cite{Veitch_2014}, that lies directly above one of the facets of the Wigner polytope\footnote{See \cite{torpedo-game} for another application of the strange state.}. Much like Bravyi and Kitaev's qubit $\ket{T}$-state, distillation of the qutrit strange state is poorly understood.  As discussed in \cite{jain2020qutrit}, noisy $\ket{S}$ states can be twirled via Clifford unitaries to lie on a line connecting a pure $\ket{S}$ state to the maximally mixed state:
\begin{equation}
    \hat{\rho}(\epsilon)=(1-\epsilon) \ket{S}\bra{S} + \epsilon \frac{1}{3}{\hat{I}_{3\times 3}}, \label{epsilon-definition}
\end{equation}
with all noise parameterized by a single parameter, $\epsilon$. Any state $\hat{\rho}(\epsilon)$ for $\epsilon <3/4$ lies outside the Wigner polytope and exhibits contextuality with respect to stabilizer measurements. We then ask whether or not there exists a family of $n$-to-$1$ magic state distillation routines that distill the strange state with a threshold approaching $\epsilon = 3/4$ as $n \to \infty$? \footnote{An additional consideration is the increase in overhead cost of distillation as $n \to \infty$.}

At the time \cite{nature} was published, and for several years thereafter, while some qutrit and qudit magic state distillation routines had been proposed \cite{ACB, CampbellAnwarBrowne, campbellEnhanced, HowardSmallCodes, DawkinsHoward}, no magic state distillation routine that distilled the Howard van Dam strange state was known. It was later discovered that an 11-qutrit CSS code based on the ternary Golay code can distill the $\ket{S}$ state, with a threshold of $ \epsilon_* = 0.38$ \cite{2020golay}. Do there exist any other qutrit stabilizer codes that distill the strange state? If so, how do their thresholds compare to that of the 11-qutrit Golay code?\footnote{A previous claim in \cite{SharmaGarani2024} suggested the existence of $[[13,1]]_3$ and $[[29,1]]_3$ codes that distill the qutrit strange state with thresholds very close to the theoretical limit of $3/4$. However, an erratum \cite{erratum} has since been published clarifying that these codes do not, in fact, distill the strange state at all. See Appendix \ref{Appendix-B}. Our independent analysis via the weight-enumerator formalism developed in this paper confirms that the 11-qutrit Golay code remains the highest-threshold distillation routine for the strange state currently known. }
In this paper, we carry out a computational search over reasonably small qutrit error-correcting codes to help answer these questions.  

One of the difficulties in finding codes that distill the strange state is that computing the performance of a distillation routine for most magic states, such as Bravyi and Kitaev's $\ket{T}$ state, requires somewhat \textit{ad hoc} methods, e.g., \cite{MSD, ACB}. This is to be contrasted with Bravyi and Kitaev's  $\ket{H}$ state \cite{MSD} and its qudit generalizations \cite{CampbellAnwarBrowne, HowardVala}, where the theory of magic state distillation is much better understood \cite{Bravyi_2012}, enabling 
systematic searches \cite{Nezami_2022} and more general constructions based on triorthogonal codes \cite{Haah_2017, Haah_2018, Hastings_2018, CampbellAnwarBrowne, campbell2014enhanced, Krishna_2019, Prakash:2024env}. But states distilled by triorthogonal codes lie above a hyperedge of the Wigner polytope and therefore demonstrating the existence of a tight distillation routine for such states would not 
demonstrate that all states outside the Wigner polytope can be 
distilled. 

One of the main results of this paper, which enables systematic searches over codes with as large as $23$ qutrits, is a simple theorem connecting the performance of a stabilizer code for qudit magic state distillation to its complete weight enumerator. For distillation routines for the qutrit strange state, this formula simplifies drastically and depends only on the simple weight enumerator of the stabilizer code. These results drastically simplify the problem of studying magic state distillation, and allow for a systematic search much larger than those previously carried out in the literature for qubits.

Our search relies on the existing classifications of qutrit error-correcting codes in the literature -- namely, the classification of qutrit stabilizer states in \cite{Danielsen2009, Danielsen2012} and a classification of self-orthogonal classical ternary codes available on \cite{munemasa_codes_website}. We carried out a search over all $[[n,1]]_3$ stabilizer codes with $n \leq 9$, and a search over all $[[11,1]]_3$ stabilizer codes that can be obtained from a $[[12,0,6]]_3$ stabilizer state. For such codes, we demand transversality of a particular single-qutrit gate (the square of the qutrit Hadamard gate), which allows us to restrict our search to projection onto the trivial syndrome of each stabilizer code. We also searched over all $[[n,1]]_3$ for odd $n \leq 23$ that possess a complete set of transversal single-qudit Clifford gates, which are necessarily CSS codes constructed from two copies of a maximal self-orthogonal ternary code.

We found that none of the codes we searched with $n<23$ could distill the $\ket{S}$ state with better-than-linear\footnote{We also found a few 9 and 11 qutrit codes that could distill the state with linear noise suppression.} noise suppression, other than the 11-qutrit Golay code of \cite{2020golay}. However, for $n=23$, we found over $600$ CSS codes that could distill the strange state with cubic noise suppression -- which is approximately $1/3$ of all the codes we could construct from the ternary self-orthogonal codes listed in \cite{munemasa_codes_website} -- suggesting that for large codes, magic state distillation is somewhat generic. None of these $23$-qutrit codes, however, had a threshold that exceeds that of the 11-qutrit Golay code.

To our knowledge, no systematic searches for qutrit 
distillation routines have appeared in the literature. Indeed, even for 
the qubit $\ket{T}$ state very few systematic searches have been carried out to date; the only examples we are aware of are \cite{MSD, reichardt2005quantum} who appear to have studied only a handful of codes, and make no claims of an exhaustive search over codes smaller than a given size, and \cite{rall2017signed} only searched over qubit codes of length $n \leq 7$. We wish to emphasize that the computational search we present in this paper appears to be the largest search possible with
 present-day technology -- ternary self-orthogonal codes with more than $23$ trits
 have not yet been classified in the coding theory literature; and, 
moreover, computing the weight enumerator of any one such code with ($n=29$) takes 6-12 hours of computational time. We expect that extending this search further would require months of computational time, at the least.

We should caution the reader that the new distillation routines we find here are mainly of theoretical interest. The success probabilities are quite low, and far better yields are obtained via triorthogonal codes \cite{Bravyi_2012, Haah_2017, Haah_2018, Hastings_2018} (see \cite{CampbellAnwarBrowne, campbell2014enhanced, Krishna_2019, Prakash:2024env} for constructions of qutrit and qudit triorthogonal codes). Nevertheless, the CSS codes we study have a complete set of transversal Clifford gates, and may turn out to be useful for fault-tolerant quantum computation in other settings.  

Our paper is organized as follows. In section \ref{sec:prelim} we briefly review some background material. In section \ref{sec:weight-enumerators} we derive a relation between weight-enumerators and the performance of magic state distillation routines. In section \ref{sec:computational-search} we describe our search space and the results. In section \ref{sec:discussion} we conclude with some brief discussion. In Appendix \ref{appendix-A} we present some useful lemmas that describe the action of stabilizer projectors on discrete phase space. In Appendix \ref{Appendix-B} we present two codes that do not distill the strange state.

\section{Preliminaries}
\label{sec:prelim}

In this section we review many basic results concerning the stabilizer formalism for qudits of odd-prime dimension $p$. We present the Heisenberg-Weyl displacement group, qudit stabilizer codes and discrete Wigner functions. The reader is directed to \cite{Wootters1987, Gottesman1999, Gibbons:2004dij, appleby2008spectra, Gross} for more details. A recent (unpublished draft) textbook which covers some of this material is \cite{Gottesman:QECC2024}.

\subsection{Heisenberg-Weyl operators and the Clifford group}

Following \cite{nature}, we will reserve the term qudits to refer to quantum systems of odd prime dimension $p$. For qudits \cite{Gottesman1999}, the Pauli group is also known as the Heisenberg-Weyl displacement group. It is defined to be generated by
$$ \hat{X} = \sum_k \ket{k+1}\bra{k}, \quad \hat{Z} = \sum_k \omega^k \ket{k}\bra{k},$$ and multiplication by $\omega= e^{2\pi i/p}$. {The operators $\hat{X}$ and $\hat{Z}$ are used to define Heisenberg-Weyl displacement operators \cite{Gross} as follows, using the conventions of \cite{nature},} 
\begin{equation}
\hat{D}(u,v) = {\omega^{2^{-1} u v}\hat{X}^u \hat{Z}^v}. \label{Heisenberg-Weyl}
\end{equation}
Heisenberg-Weyl displacement operators {acting on $n$ qudits} are denoted as
\begin{equation}
    \hat{D}(\vec{u},\vec{v}) = \hat{D}(u_1,v_1) \otimes \hat{D}(u_2,v_2) \otimes \ldots \otimes \hat{D}(u_n,v_n), \label{multi-qudit-operator}
\end{equation}
where $\vec{u}=(u_1, u_2, \ldots, u_n)$ and $\vec{v}=(v_1, v_2, \ldots, v_n)$. It is convenient to combine $\vec{u}$ and $\vec{v}$ into a \textit{symplectic vector} $\chi=(\vec{u},\vec{v})$, and write $\hat{D}(\chi) =\hat{D}(\vec{u},\vec{v})$. The \textit{Hamming weight} of a multi-qudit Heisenberg-Weyl displacement operator defined by a symplectic vector $\chi$, is defined as the number of entries such that $\chi_i = (u_i,v_i) \neq (0,0)$, just as for multi-qubit Pauli-operators.

Multiplication of Heisenberg-Weyl operators corresponds to addition of symplectic vectors, with the possible introduction of an additional overall phase,
\begin{equation} 
\hat{D}({\chi})\hat{D}(\chi') = \omega^{2^{-1}  [\chi, \chi'] } {\hat{D}({\chi+\chi'}).}\label{eq:overall-phase}
\end{equation}
Here {$2^{-1}$ is the inverse of 2 in the field $\mathbb Z_p$} and
\begin{equation}
    [\chi, \chi']= \vec{u} \cdot \vec{v}'- \vec{u}' \cdot \vec{v}
\end{equation} 
is the symplectic inner product. For a pair of commuting Heisenberg-Weyl operators $[\chi,\chi']$ vanishes. Because of equation \eqref{eq:overall-phase}, the Heisenberg-Weyl group includes operators with additional overall phases, such as $\omega^a \hat{D}(\vec{u},\vec{v})$, with $a \in \mathbb Z_p$.  

The correspondence between symplectic vectors and Heisenberg-Weyl displacement operators plays an important role in this paper, so let us discuss it in more detail. An element of the $n$-qudit Heisenberg-Weyl displacement group is uniquely specified by a symplectic vector $\chi=(\vec{u},\vec{v}) \in \mathbb Z_p^{2n}$, and a phase $\omega^a$. We define \textit{phase-free} Heisenberg-Weyl displacement operators to be operators of the form $D(\vec{u},\vec{v})$ in equation \eqref{Heisenberg-Weyl}, without any overall phase. The set of all $n$-qudit phase-free Heisenberg-Weyl displacement operators is in one-to-one correspondence with the set of symplectic vectors $\chi$. While symplectic vectors form a group under addition, the phase-free Heisenberg-Weyl operators do not form a group, since multiplication of two operators can induce an overall phase, as per equation \eqref{eq:overall-phase}. However, a set of \textit{mutually commuting} phase-free Heisenberg-Weyl operators generates a subgroup of the Heisenberg-Weyl displacement group consisting entirely of phase-free Heisenberg-Weyl operators, and is isomorphic to a subspace of symplectic vectors. 

A unitary operator $\hat{C}$ is said to be a Clifford operator if it maps Heisenberg-Weyl displacement operators to Heisenberg-Weyl displacement operators under conjugation:
\begin{equation}
    C D(\chi) C^\dagger = \omega^a D(\chi').
\end{equation}
The set of $n$-qudit Clifford unitaries form a group. The single-qudit Clifford group is generated by two operators, $\hat{H}$, and $\hat{S}$, defined as,
\begin{equation}
    \hat S = \sum_{j=0}^{p-1} \omega^{2^{-1}j(j+1)}\ket{j}\bra{j}, \quad \hat H = \frac{1}{\sqrt{p}}\sum_{j=0}^{p-1} \sum_{k=0}^{p-1} \omega^{jk}\ket{j}\bra{k}.
\end{equation}
Many useful properties of the Clifford group for qudits of odd prime dimension are given in \cite{appleby2008spectra}. In particular, up to phases, any single-qudit Clifford operator can be written as a symplectic rotation, followed by a Heisenberg-Weyl displacement operator, $C =\hat{D}(\chi) \hat{V}_F$. Symplectic rotations are operators $\hat{V}_F$ that satisfy $\hat V_F^{-1}\hat D(
\chi) \hat V_F=\hat D(\chi')$, with 
\begin{equation}
    \chi' = F \chi, \quad F=\begin{pmatrix} a & b \\ c & d \end{pmatrix}.
\end{equation}
$F$ is a linear transformation that preserves the symplectic inner-product. For a single qudit,  $F \in SL(2,\mathbb Z_p)$. Symplectic rotations are generated by $\hat{H}$ and $\hat{Z}\hat{S}$. To see this, note that, $\hat{H}^{-1} \hat{D}(u,v) \hat{H} = D(v,-u)$, and $\hat{S}^{-1}\hat{Z}^{-1} \hat{D}(u,v) \hat{Z}\hat{S}=\hat{D}{(u,v-u)}$. $\hat{H}$ and $\hat{Z}\hat{S}$ thus correspond to $\begin{pmatrix} 0 & -1 \\ 1 & 0 \end{pmatrix}$ and $\begin{pmatrix} 1 & -1 \\ 0 & 1 \end{pmatrix}$; together these generate all of $SL(2,\mathbb Z_p)$. Observe that, for qudits $\hat{H}^2 \neq \hat{I}_{3\times 3}$, and instead, $\hat{H}^2$ acts as,
\begin{equation}
    \hat{H}^2\hat{D}(u,v) \hat{H}^{-2}=\hat{D}(-u,-v)=\hat{D}(u,v)^{-1}. \label{h2-action}
\end{equation}

\subsection{Discrete Wigner functions}

{In this paper, we will make extensive use of a discrete phase space formalism for qudits. This was first formulated in \cite{Wootters1987, Gibbons:2004dij} and played a central role in \cite{nature}.  \cite{Gross, Veitch_2012, Veitch_2014} and subsequently \cite{wang2018efficiently, majorization} used this formalism to define the resource theories of magic. Many examples of discrete Wigner functions for qudits are given in \cite{jain2020qutrit}, and it was also used in \cite{prakash2020contextual, 2020golay}. Here, we provide a very brief review of the essential features of this formalism.

In essence, the discrete Wigner function is a convenient way to represent single-qudit and multi-qudit density matrices for qudits of odd-prime dimension.} It is constructed using phase-point operators, which, for a {single}-qudit, are defined using the Heisenberg-Weyl displacement operators, as,
\begin{equation}
     \hat{A}(0,0) = \frac{1}{p} {\sum_{u=0}^{p-1} \sum_{v=0}^{p-1}} \hat{D}(u,v), \quad  \hat{A}(u,v) =     \hat{D}(u,v) \hat{A}(0,0)  \hat{D}(u,v)^\dagger.
\end{equation}
{Using the fact that $\hat{A}(0,0)=\frac{1}{p}\hat{H}^2$, one can show that,
\begin{equation}
    \tr A(u,v)A(u',v') = \delta_{u,u'}\delta_{v,v'}.
\end{equation}}

Multi-qudit phase point operators are defined as 
\begin{equation}
    \hat{A}(\vec{u},\vec{v})=\bigotimes_i \hat{A}(u_i,v_i).
\end{equation}
{The phase-point operators form a basis for qudit density matrices, normalized so that $\tr A(\vec{u},\vec{v})=1$. Any $n$-qudit density matrix $\hat{\rho}^{(n)}$ can be written as a linear combination of the phase point operators with real, but possibly negative, coefficients. These coefficients define the discrete Wigner function $W( \hat{\rho}^{(n)}; \vec{u},\vec{v})$ of the qudit state.} Explicitly, for a single-qudit state $\hat{\rho}$,
\begin{equation}
    \hat{\rho} = \sum_{u=0}^{p-1}\sum_{v=0}^{p-1} W( \hat{\rho}; u,v)\hat{A}(u,v), \quad W( \hat{\rho}; u, v) = \frac{1}{p} \tr  \left(\hat{\rho} \hat{A}(u,v)\right) \label{eq:discrete-wigner-def}
\end{equation}
Note that $\tr  \hat{\rho} = \sum_{u,v} W( \hat{\rho}; u,v)$. {As explained in \cite{Wootters1987, Gibbons:2004dij, Gross}, the discrete Wigner function defines a quasi-probability distribution for stabilizer measurements, much like the original continuous Wigner function \cite{Wigner1932}.}  

The Clifford group acts covariantly on the discrete phase space: general Clifford transformations $\hat{D}(\chi)\hat{V}_F$ act as symplectic rotations followed by translations \cite{Gross, appleby2008spectra, Veitch_2012}:
\begin{equation}
    \hat{D}(\chi')\hat{V}_F \hat A(\chi) (\hat{D}(\chi')\hat{V}_F)^{-1}=\hat A(F\chi +\chi').
\end{equation}
This allows one to use the discrete Wigner function to define an efficient classical simulation for Clifford unitaries and stabilizer measurements acting on qudit states with non-negative Wigner functions \cite{Veitch_2012}. The only pure states with non-negative Wigner functions are stabilizer states \cite{Gross}; however, the set of mixed states with non-negative Wigner function is larger than the set of mixtures of stabilizer states \cite{Veitch_2012}. This means that this simulation of \cite{Veitch_2012} is more powerful than the Gottesman-Knill theorem \cite{GottesmanKnillTheorem}, which can be directly generalized to qudits. The set of mixed states with non-negative Wigner function forms a convex polytope known as the Wigner polytope \cite{Veitch_2012}; because such states can be classically simulated, they must be useless for magic state distillation. \cite{nature} gave a more foundational interpretation of the Wigner polytope, by showing that any state with negative Wigner function exhibits contextuality with respect to stabilizer measurements, and vice-versa.

As an example, let us compute the discrete Wigner function for the qutrit strange state, which is the magic state of primary interest in this work. The qutrit strange state is given by \cite{HowardVanDam, Veitch_2012, jain2020qutrit} 
\begin{equation}
    \ket{S}= \frac{1}{\sqrt{2}} \left (\ket{1}- \ket{2} \right).
\end{equation}
Using the definitions above, its discrete Wigner function can be computed to be, \begin{equation}
    W(\ket{S}\bra{S}; u,v) = \begin{cases}\frac{1}{3} & (u,v)=(0,0) \\ \frac{1}{6} & (u,v) \neq (0,0) \end{cases} \quad = \quad \begin{pmatrix} 1/6 & 1/6 & 1/6 \\ 1/6 & 1/6 & 1/6 \\ -1/3 & 1/6 & 1/6 \end{pmatrix}. \label{wigner-function-pure-strange}
\end{equation}
It is self-evident from the form of this discrete Wigner function that $\ket{S}$ is an eigenvector of all symplectic rotations, i.e., Clifford unitaries of the form $\hat{V}_F$. Alternatively, recalling that symplectic rotations are generated by $\hat{Z}\hat{S}$ and $\hat{H}$, one can check that $\hat{H}\ket{S}=i \ket{S}$ and $\hat{Z}\hat{S} \ket{S}=\omega^2 \ket{S}$. This is discussed in much more detail in \cite{jain2020qutrit}. We will take advantage of the particularly simple form of this Wigner function in what follows.

Let us next discuss noisy qutrit strange states. A generic noisy magic state is described by a qudit density matrix that requires $p^2-1$ real parameters to describe. To simplify the analysis of noise, \cite{MSD} introduced the idea of \textit{twirling}, which reduces the number of parameters needed to describe a noisy magic state. The procedure involves applying a randomly-chosen element from a subgroup of the Clifford group to the state, effectively averaging its density matrix over the orbit of that subgroup. For this to be a useful simplification (i.e., to average the noise without affecting the state), the pure magic state must be invariant under the chosen subgroup. 

After twirling by the subgroup of the qutrit Clifford group consisting of symplectic rotations, as described in \cite{jain2020qutrit}, noisy $\ket{S}$ states are described by the one-parameter family of density matrices:
\begin{equation}
    \hat{\rho}_S(\epsilon)=(1-\epsilon) \ket{S}\bra{S} + \epsilon \frac{1}{3}\hat{I}_{3\times3}. \label{mixed}
\end{equation}
We require $0 \leq \epsilon \leq 3/2$ for equation \eqref{mixed} to describe a valid density matrix.  Using equation \eqref{eq:discrete-wigner-def}, we compute the discrete Wigner function of $\hat{\rho}_S(\epsilon)$ to be:
\begin{equation}
    W(\hat{\rho}_\epsilon; u,v) = \begin{cases} x & (u,v)=(0,0) \\ y & (u,v) \neq (0,0) \end{cases} \quad = \quad \begin{pmatrix} y & y & y \\ y & y & y \\ x & y & y \end{pmatrix}. \label{wigner-function}
\end{equation}
{where $x$ and $y$ are given by,}
\begin{eqnarray}
3x & = & -1+4\frac{\epsilon}{3} \label{xeps} \\
6y & = & 1-\frac{\epsilon}{3} \label{yeps}. 
\end{eqnarray}
Note that $8y+x=1$, {as required by normalization of the Wigner function.} 

If contextuality is sufficient for universal quantum computing, we should be able to distill pure strange states from any strange state of the form \eqref{wigner-function}, with $x<0$. Moreover, \textit{any} qutrit state $\rho_{\rm in}$ outside the Wigner polytope can be put into the form given by \eqref{wigner-function}, with $x<0$, using only Clifford unitaries, as follows. Any state that lies outside the Wigner polytope has a negative entry: $W(\rho_{\rm in};i,j)<0$.  Acting on $\rho$ with $D(-i,-j)$ one obtains $\hat{\rho}'$ with $W(\hat{\rho}_{\rm in}';0,0)<0$. Twirling with symplectic rotations preserves $W(\hat{\rho}'; 0,0)$ so $x$ remains negative after twirling. Therefore, demonstrating the existence of a magic state distillation routine that distills pure magic states from twirled strange states with the optimum threshold demonstrates that any state outside the Wigner polytope can be distilled.

\subsection{Qudit stabilizer codes}
Stabilizer codes play a crucial role in magic state distillation. Qubit stabilizer codes are discussed in many textbooks \cite{NielsenChuang2010}, and in \cite{PhysRevA.57.127}. While qudit stabilizer codes are less well-studied, some of their basic properties are worked out in \cite{Gottesman1999, ketkar2006nonbinary}, and a nice introduction can be found in \cite{Gottesman:QECC2024}. Here, we review the essential features that we will use, and establish some conventions.

The codespace of an $[[n,k]]_p$ stabilizer code is defined by the simultaneous eigenspace of $n-k$ independent commuting $n$-qudit Heisenberg-Weyl displacement operators, and has dimension $p^k$. We adopt the convention in this paper, that each of these $n-k$ commuting operators are phase-free. To completely specify the eigenspace, we also have to specify the eigenvalues of each stabilizer, which are of the form $\omega^a$, where $a \in \mathbb Z_p$. We will sometimes refer to these eigenvalues as the \textit{syndrome} of the eigenspace. In this paper, we will {usually} restrict our attention to codes defined by the $\omega^0=1$ eigenspace of each operator, i.e., the trivial syndrome. 

The set of commuting operators $\{ \hat{D}(\vec{u}_1, \vec{v}_1), \ldots, \hat{D}(\vec{u}_{n-k}, \vec{v}_{n-k}) \}$ can be specified by a symplectic matrix
\begin{equation}
H = \left( \begin{array}{c|c}
    \vec{u}_{1} & \vec{v}_{1} \\
    \vec{u}_{2} & \vec{v}_{2} \\
    \vdots & \vdots \\
    \vec{u}_{n-k} & \vec{v}_{n-k} 
    \end{array}\right),  \label{stabilizer-generator-matrix}
\end{equation}
such that the symplectic inner product of any two rows vanishes. These operators generate an abelian group of order $p^{n-k}$, which we refer to as the stabilizer group $\mathcal S$. Because $\mathcal S$ is abelian, there are no overall phases arising from multiplication and $\mathcal S$ consists exclusively of phase-free Heisenberg-Weyl displacement operators, each of the form $\hat{D}(\chi)$ where $\chi$ is in the row-span of $H$. The group $\mathcal S$ of $n$-qudit commuting phase-free Heisenberg-Weyl displacement operators is isomorphic to the subspace of symplectic vectors in $\mathbb Z_p^{2n}$ spanned by the rows of $H$, where multiplication of Heisenberg-Weyl operators corresponds to addition of symplectic vectors. We will use $\mathcal S$ to refer, interchangeably, to this subspace of symplectic vectors or the group of commuting phase-free Heisenberg-Weyl displacement operators.

{We will frequently make use of the correspondence between stabilizer codes and additive codes over $GF(p^2)$. This correspondence is well-known for qubits \cite{calderbank1998quantum}, and its extension to qudits, which is straightforward, is derived in \cite{ ketkar2006nonbinary}. A more pedagogical discussion appears in  \cite{Gottesman:QECC2024}. Essentially, symplectic vectors in $\mathbb Z_p^{2n}$ can be thought of as vectors in $GF(p^2)^n$, and linear subspaces of $\mathbb Z_p^{2n}$ correspond to subsets of $GF(p^2)^n$ that are closed under addition. The symplectic inner product translates into the Hermitian inner product. Therefore, a stabilizer code can equivalently be thought of as an additive classical code over $GF(p^2)$, that is self-orthogonal under the Hermitian inner product. In particular, it is conventional to consider $H$, in equation \eqref{stabilizer-generator-matrix}, to be the generator matrix for the classical additive code over $GF(p^2)$. As such an $[[n,k]]_p$ stabilizer code corresponds to an additive $(n, p^{n-k})_{GF(p^2)}$ code, using the notation of \cite{calderbank1998quantum}.}

{We use the notation $N(\mathcal S)$ to denote all $n$-qudit Heisenberg-Weyl displacement operators that commute with $\mathcal S$. We will also use the notation $\mathcal S^\perp$ to denote the subspace of symplectic vectors orthogonal to the self-orthogonal subspace of symplectic vectors $\mathcal S$. Clearly, $\mathcal S \subseteq N(\mathcal S)$ and $\mathcal S \subset \mathcal S^\perp$. Unless $k=0$, $N(\mathcal S)$ will be a non-abelian group, and multiplication of two different elements may give rise to Heisenberg-Weyl operators with overall phases. As such, all the operators in $N(\mathcal S)$ will not be phase-free, and there is not a one-to-one correspondence between $N(\mathcal S)$ and $\mathcal S^\perp$. However, we can define the subset $N(\mathcal S)^*$ of phase-free operators in $N(\mathcal S)$, which is in one-to-one correspondence with $\mathcal S^\perp$.  We use the notation $\mathcal S^\perp$ and $N(\mathcal S)^*$ interchangeably.} 

{Any symplectic vector $L \in \mathcal S^\perp$ corresponds to a logical operator for the stabilizer code. This logical operator is the identity operator if $L \in \mathcal S$, and is non-trivial otherwise. We define the \textit{coset} $[L]$ of $\mathcal S$ in $\mathcal S^\perp$ to consist of all representatives of the logical operator corresponding to $L$; if $L \in \mathcal S$, the coset is simply the trivial coset $\mathcal S$, else it is non-trivial. All operators in the same coset commute with each other, but operators from different non-trivial cosets do not commute with each other.  Viewing $\mathcal S$ as a subspace of symplectic vectors, the coset $[L]$ of $\mathcal S$ in $\mathcal S^\perp$ is the set of symplectic vectors of the form $\mathcal S + L = \{ L + M | M \in \mathcal S\}$.} 

{The smallest Hamming weight of a non-trivial logical operator $L \in \mathcal S^\perp/\mathcal S$ is defined to be the \textit{distance} $d$ of the stabilizer code. Computing the distance of a generic stabilizer code is computationally non-trivial; when the distance of an $[[n,k]]_p$ stabilizer code is known, it is usually referred to as an $[[n,k,d]]_p$ stabilizer code. In the special case $k=0$, the stabilizer code has no non-trivial logical operators, and the distance of a stabilizer code is instead conventionally defined as the minimum Hamming weight of any stabilizer \cite{calderbank1998quantum, Gottesman:QECC2024}.}

Let $\mathcal S$ denote the stabilizer group of an $[[n,k,d]]_p$ stabilizer code. Let $\{M_1, \ldots, M_{n-k}\}$ be a set of generators for a stabilizer code. The projector onto the $+1$ eigenspace of $M_i$ is given by $$\Pi_{M_i} = \frac{1}{p}\left(1+M_i + M_i^2 + \ldots M_i^{p-1}\right).$$ To see that this is true, note that the eigenvalues of $M_i$ are $\omega^a$.  If $M_i \ket{\psi}=\omega^a \ket{\psi}$, then, 
\begin{equation}
\Pi_{M_i}\ket{\psi}=\frac{1}{p}(1+\omega^a+\omega^{2a} + \ldots + \omega^{(p-1)a})\ket{\psi}=
\begin{cases} 
1\ket{\psi} & a =0 \\ 0 & a \neq 0.
\end{cases}
\end{equation}  The projector onto the codespace of $\mathcal S$ with trivial syndrome can be written explicitly as,
\begin{equation}
    \hat{\Pi}^0_{\mathcal{S}} = {\prod_{i=1}^{n-k} \frac{1}{p}\left(1+\hat{M}_i + \hat{M}_i^2 + \ldots \hat{M}_i^{p-1}\right) =}\frac{1}{p^{n-k}} \sum_{ \hat{M} \in \mathcal{S} } \hat{M}. \label{trivial-projector}
\end{equation}

We say a Clifford unitary is \textit{transversal} if it commutes with $\Pi^{\vec{s}}_{\mathcal S}$. Depending on context, we may, implicitly, also demand that it acts as the logical operator $\bar{C}$ or $\bar{C}^{-1}$.

\subsection{Weight enumerators}
Here, we define the weight enumerator of a stabilizer code. The most convenient way to do this is to make use of the correspondence between classical codes over $GF(p^2)$ and stabilizer codes. Simple and complete weight enumerators for classical codes over an arbitrary finite field are defined in \cite{MacWilliamsSloane}. {Weight enumerators for quantum error-correcting codes were defined in more generality in \cite{shor1997quantum} -- for stabilizer codes, the definitions of \cite{shor1997quantum} coincide with the classical definitions of the complete and simple weight enumerators for error-correcting codes over the finite field $GF(p^2)$. Our presentation below differs from that in \cite{shor1997quantum}, and is essentially a translation of the definition of weight-enumerators given in \cite{MacWilliamsSloane} for classical codes over $GF(p^2)$ to the language of stabilizer codes.}

Let $\tilde{ \mathcal S}$ be either a stabilizer code $\mathcal S$, {or one of its cosets $[L]=\mathcal S+L$ in $\mathcal S^\perp$.} We define the \textit{complete weight enumerator} of $\tilde{\mathcal S}$ to be a function of $p^2$ formal variables $\{y_{\alpha \beta}\}$, for $\alpha, \beta \in \mathbb Z_p$, defined as follows:
\begin{equation}
 w(\tilde{\mathcal S}; ~ {   \{y_{\alpha \beta}\}} ) = \sum_{(\vec{x}|\vec{z}) \in \tilde{ \mathcal S}} ~ \prod_{i=1}^n {y_{x_i,z_i}}. \label{complete-weight-enumerator}
\end{equation}
One can think of this map as arising from a formal operation $\mathcal F$, which is defined recursively via the rules \begin{equation}
    \mathcal F[A \otimes B]=\mathcal F[A] \cdot  \mathcal F[B], \quad \mathcal F[A + B]=\mathcal F[A] + \mathcal F[B], \quad \mathcal F[\omega^a A]=\omega^a \mathcal F[A],
\end{equation}
and the base case, $\mathcal F\left[\hat D(u,v)\right]=y_{u,v}$. 

Then $w(\tilde{\mathcal S}; {   y_{ij}} ) = p^{n-k}\mathcal F\left[ \hat{\Pi}^0_{\mathcal{S}} \right]$.

The \textit{simple weight enumerator} of $\tilde{ \mathcal S}$ is a function of two formal variables $x$ and $y$, defined as follows:
\begin{equation}
    w(\tilde{ \mathcal S}; x, y) =  w(\tilde{ \mathcal S}; ~\{y_{\alpha \beta}(x, y)\} ).  \label{simple-weight-enumerator}
\end{equation} 
where 
\begin{equation}
    y_{\alpha \beta}(x,y) = \begin{cases} x & ({\alpha,\beta})=(0,0) \\
    y & ({\alpha,\beta}) \neq (0,0). \end{cases}
\end{equation}
{This has the interpretation as a generating function for the Hamming weights of the stabilizers in $\mathcal S$.} If we further set $x=y=1$, then the simple weight enumerator becomes equal to $|\tilde{ \mathcal S}|$.

A MacWilliams identity relates $w(\mathcal S;1,z)$ to $w(\mathcal S^\perp; 1,z)$ \cite{MacWilliamsSloane,shor1997quantum, ketkar2006nonbinary}. This is:
\begin{equation}
    w(\mathcal S^\perp;1,z) = \frac{(1+(p^2-1)z)^n}{p^{n-k}} w\left(\mathcal S;1,\frac{1-z}{1+(p^2-1)z} \right).
\end{equation}
It is conventional to define $A_{\mathcal S}(z)=w(\mathcal S;1,z)$ and $B_{\mathcal S}(z)=w(\mathcal S^\perp; 1,z)$. 
{$B_{\mathcal S}(z)-A_{\mathcal S}(z)$} is a polynomial in $z$ with non-negative coefficients and is the simple weight enumerator of the set of all logical operators for the stabilizer code $\mathcal S$. The lowest power of $z$ that appears in {$B_{\mathcal S}(z)-A_{\mathcal S}(z)$} is $z^d$, where $d$ is the distance of the stabilizer code.

\section{Distillation and weight enumerators}
\label{sec:weight-enumerators}
In the magic state model of fault-tolerant quantum computing \cite{MSD, Knill2004QuantumCW}, one begins with a quantum computer that can initialize qudits in the computational basis, perform Clifford unitaries, and carry out stabilizer measurements. These operations are assumed to be noise-free. To obtain universal quantum computing, we supplement this quantum computer with the ability to initialize ancilla qudits in certain non-stabilizer states known as magic states, which are noisy. A magic state distillation protocol is a way to distill an arbitrarily pure magic state from many noisy magic states using only Clifford unitaries and stabilizer measurements. For the purposes of determining the best attainable threshold \cite{campbell2009structure}, any magic state distillation protocol for the qutrit strange state can be thought of as a procedure that projects $n$ noisy qudits onto the codespace of an $[[n,k]]_p$ stabilizer code -- if the projection is successful, one decodes the resulting qudits to obtain $k$ hopefully-less-noisy magic states.\footnote{In some cases, namely, distillation of Bravyi and Kitaev's $\ket{H}$ state via the $[[15,1,3]]$ code of \cite{MSD} and its qudit analogues \cite{CampbellAnwarBrowne}, error-correction is also possible (although not necessarily advantageous) prior to decoding. However, this is not possible for distillation routines for the qutrit strange state.}

\subsection{A general formulation in terms of complete weight enumerators}
Here we provide a general formulation for qudit magic state distillation in the language of complete weight enumerators. These results follow in part from the formulation in \cite{prakash2020contextual}. Let us also mention that a similar formulation in terms of signed-weight enumerators for qubit magic state distillation was given in \cite{rall2017signed}. 

In magic state distillation, we first project $n$ noisy input states $\rho_{\rm in}$ onto the codespace of an $[[n,k]]_p$ $\mathcal S$ to obtain a new $k$-qudit output state $\hat{\rho}' = f_{MSD}(\rho_{\rm in})$. The procedure succeeds with probability $\nu$. Our main result is the following theorem that expresses both $\nu$ and the discrete Wigner function of the $\hat{\rho}'$ in terms of the complete weight enumerators of $\mathcal S$ and its cosets in $\mathcal S^\perp$.

\newtheorem{lemma}{Lemma}
\begin{theorem}
    Let $\hat \rho$ be a  single qudit mixed state described by the Wigner function $W(\hat \rho; \alpha, \beta)$ and let $\mathcal S$ be an $[[n,k]]_p$ stabilizer code with trivial syndrome. 
    \begin{itemize}
        \item[(a)] The probability for successful projection onto the eigenspace of $\mathcal S$ with trivial syndrome is given by the \textbf{complete weight enumerator} of $\mathcal S^\perp$, $w( \mathcal S^\perp; \{ W(\hat \rho;\alpha,\beta) \})$, with the formal variables $y_{\alpha \beta}$ in the complete {w}eight enumerator replaced by the entries of the Wigner function of $\rho$:
\begin{equation}
    \tr \left( \hat{\Pi}^0_{\mathcal S} {\rho^{\otimes n} }\right) = w( \mathcal S^\perp; ~\{ W(\hat \rho;\alpha,\beta) \}) = \sum_{(\vec{u}|\vec{v}) \in \mathcal S^\perp} \prod_{i=1}^n W(\hat \rho; u_i, v_i).
\end{equation}
 
    \item[(b)] Let $\bar{D}(\vec{u},\vec{v})$ be logical operators for the stabilizer code, with $\vec{u}, \vec{v} \in \mathbb Z_p^k$. If the projection from part (a) is successful, the output state $\hat \rho^{(n)}_{\rm out}$ corresponds to the logical state $\hat{\rho}'$, whose discrete Wigner function is given by,
    \begin{equation}
    W(\hat \rho';\vec{u},\vec{v}) = \frac{1}{\nu} w\left( [\bar{D}(\vec{u},\vec{v})];~ \{ W(\hat{\rho};\alpha,\beta) \}\right). \label{msd-2}
\end{equation}
   \end{itemize}  
\label{theorem-1}
\end{theorem}

\begin{proof}[Proof of part (a)]

The $n$-qudit Wigner function for $\rho^{\otimes n}$ is given by, 
\begin{equation}
    W^{(n)}_{\rm in}(\hat \rho^{\otimes n}; \vec{u},\vec{v}) = \prod_i W_{}(\hat{\rho}; u_i,v_i), \label{W-in}
\end{equation}
and we can write 
\begin{eqnarray}
    \rho^{\otimes n} = \sum_{\vec{u} \in \mathbb Z_p^n}\sum_{\vec{v} \in \mathbb Z_p^n} \prod_i W_{}(\hat{\rho}; u_i,v_i) \hat{A}(\vec{u},\vec{v}).
\end{eqnarray}
The probability for successful projection onto the codespace is,
\begin{eqnarray}
    \nu & = & \tr (\hat{\Pi}^0_{\mathcal S} \hat \rho^{\otimes n}) \\
    & = & \sum_{\vec{u}, \vec{v}} \left(\prod_{i=1}^n W(u_i,v_i)\right)  \tr \left( \hat{\Pi}^0_{\mathcal S} \hat{A}(\vec{u},\vec{v}) \right),
\end{eqnarray}
where $\Pi^0_{\mathcal S}$ is the projector onto the eigenspace of $\mathcal S$ with trivial syndrome.
In Appendix \ref{appendix-A}, Lemma \ref{phase-point-lemma},  we show that 
\begin{eqnarray}
   \tr \left(  \hat{\Pi}^0_{\mathcal S} \hat{A}(\vec{u},\vec{v}) \right)
   & = & \begin{cases} 0 & (\vec{u}|\vec{v}) \notin \mathcal S^\perp \\
   1 & (\vec{u}|\vec{v}) \in \mathcal S^\perp
   \end{cases}.
\end{eqnarray}
We therefore find that,
\begin{equation}
   \nu  =  \sum_{(\vec{u}|\vec{v}) \in \mathcal S^\perp } \prod_i W(\hat{\rho}; u_i,v_i).
\end{equation}
\end{proof}

Part (a) of Theorem \ref{theorem-1} determines the probability of successfully projecting $\rho^{\otimes n}$ onto the codespace of an $[[n,k]]_p$ stabilizer code, $\mathcal S$. If the projection is successful, the resulting $n$-qudit output state $\hat \rho^{(n)}_{\rm out}$ will be given by
\begin{equation}
    \hat \rho^{(n)}_{\rm out} = \frac{1}{\nu}\left( \hat{\Pi}^0_{\mathcal S} \hat \rho^{\otimes n} \right).
\end{equation}
Part (b) tells us the logical interpretation of this state.

\begin{proof}[Proof of part (b):]
 Assuming, for simplicity, that $k=1$, so that $\hat{\rho}_{\rm out}^{(n)}$ corresponds to a single-qudit logical state we denote as $\hat{\rho}'$. Let $\bar{X}$ and $\bar{Z}$ be any representatives of the logical Pauli-operators of the code. From these, we can define logical Heisenberg-Weyl operators $\bar{D}(x,z)$ and logical phase-point operators $\bar{A}(x,z)$. Then, the discrete Wigner function of $\hat{\rho}'$ is given by,
\begin{equation}
    W(\hat{\rho}'; x,z) = \frac{1}{\nu} \tr \left( p^{-1} \bar{A}(x,z) \hat{\Pi}^0_{\mathcal S} \hat{\rho}^{\otimes n} \right).
\end{equation}
Notice that,
\begin{equation}
    p^{-1} \bar{A}(0,0) \hat{\Pi}^0_{\mathcal S} =  p^{-2}\sum_{u,v} \bar{D}(u,v) \hat{\Pi}^0_{\mathcal S} =  \frac{1}{p^{n+1}} \sum_{(\vec{u}|\vec{v}) \in \mathcal S^\perp} \hat{D}(\vec{u},\vec{v}) \equiv \hat{\Pi}^0_{\mathcal S^\perp}.
\end{equation}
Using the discrete Wigner function for $\hat \rho^{\otimes n}$ in equation \eqref{W-in}, and Lemma \ref{dual-lemma} in Appendix \ref{appendix-A}, one can show that,
\begin{equation}
    W(\hat{\rho}'; 0,0) = \frac{1}{\nu} \tr \left( \hat{\Pi}^0_{\mathcal S^\perp} \hat{\rho}^{\otimes n} \right)  = \frac{1}{\nu} w( \mathcal S; \{ W(\hat{\rho},\alpha ,\beta) \}). \label{msd-1}
\end{equation}
More generally,
\begin{equation}
    p^{-1} \bar{A}(x,z) \hat{\Pi}^0_{\mathcal S} =  p^{-2}\bar{D}(x,z)\left( \sum_{u,v} \bar{D}(u,v) \right) \bar{D}(x,z)^\dagger \hat{\Pi}^0_{\mathcal S} 
    = \bar{D}(x,z) \hat{\Pi}^0_{\mathcal S^\perp} \bar{D}(x,z)^\dagger,
\end{equation}
because $[\bar{D}(x,z)^\dagger, \hat{\Pi}^0_{\mathcal S}]=0$. Therefore
\begin{equation}
    W(\hat{\rho}';x,z) = \frac{1}{\nu} \tr \left( \bar{D}(x,z)\hat{\Pi}^0_{\mathcal S^\perp} \bar{D}(x,z)^\dagger \hat{\rho}^{\otimes n} \right) = \frac{1}{\nu} \tr \left( \hat{\Pi}^0_{\mathcal S^\perp} \bar{D}(x,z)^\dagger \hat{\rho}^{\otimes n}\bar{D}(x,z) \right)
\end{equation}
Let $D(\vec{x}_L,\vec{z}_L)$ be a representative of the logical operator $\bar{D}(x,z)$. Then the discrete Wigner function of $\bar{D}(x,z)^\dagger \hat{\rho}^{\otimes n}\bar{D}(x,z)$  is given by,
\begin{eqnarray}
    \bar{D}(x,z)^\dagger \hat{\rho}^{\otimes n}\bar{D}(x,z) & = &  \sum_{\vec{u} \in \mathbb Z_p^n}\sum_{\vec{v} \in \mathbb Z_p^n} \prod_i W_{}(\hat{\rho}; u_i,v_i) D(\vec{x}_L,\vec{z}_L)^\dagger \hat{A}(\vec{u},\vec{v}) D(\vec{x}_L,\vec{z}_L) \\ & = &
    \sum_{\vec{u} \in \mathbb Z_p^n}\sum_{\vec{v} \in \mathbb Z_p^n} \prod_i W_{}(\hat{\rho}; u_i,v_i) \hat{A}(\vec{u}-\vec{x}_L,\vec{v}-\vec{z}_L) 
\end{eqnarray}
We can then use Lemma \ref{dual-lemma} from Appendix \ref{appendix-A} to obtain, 
\begin{equation}
    W(\hat{\rho}';x,z) = \frac{1}{\nu} w( \mathcal S+(\vec{x}_L|\vec{z}_L); \{ W(\hat{\rho},\alpha ,\beta) \}) = \frac{1}{\nu} w\left( [\hat{D}(x,z)];~\{ W(\hat{\rho},\alpha ,\beta) \}\right)
\end{equation}
Recall that $[\bar{D}(x,z)]$ denotes the coset $\mathcal S+(\vec{x}_L|\vec{z}_L)$ in $\mathcal S^\perp$, and denotes a the set of all representatives of the logical operator $\bar{D}(x,z)$. 
\end{proof}

The problem of computing the output state of a general qudit magic state distillation routine defined by a stabilizer $\mathcal S$ with trivial syndrome is thus reduced to computing the complete weight enumerators of $\mathcal S^\perp$, interpreted as a classical error-correcting code over $GF(p^2)$, and its cosets. The formal variables $y_{ij}$ in these weight enumerators are replaced by the entries of the discrete Wigner function, $W(\hat{\rho};i,j)$ of the noisy input state.

{In Theorem \ref{theorem-1}, we imposed the condition that we are projecting onto the \textit{trivial} syndrome of a stabilizer code. As explained in Section \ref{sec:symmetry} below, this condition can be justified by noting that it is equivalent to demanding that $\hat{H}^2$ is a transversal gate for our code, which is a natural requirement when searching for distillation routines for magic states that are eigenvectors of $\hat{H}^2$\cite{jain2020qutrit}, including, but not limited to, the strange state. More generally, we should point out that, to our knowledge, this condition is satisfied by all qudit magic state distillation routines known to date, including distillation routines for magic states that are not eigenvectors of the strange state (e.g., those in \cite{ACB, CampbellAnwarBrowne, campbellEnhanced, Krishna_2019, Prakash:2024env}). It is straightforward to modify Theorem \ref{theorem-1} to project onto eigenspaces of stabilizer codes with non-trivial syndrome, and we sketch how to do this in Lemma \ref{phase-point-lemma-v2} in Appendix \ref{appendix-A}. }

\subsection{The strange state and simple weight enumerators}
\label{simple-weight-enumerator-section}
We now restrict our attention to the special case where $\hat{\rho}=\hat{\rho}_S(\epsilon)$ is a twirled qutrit strange state $\ket{S}$. 
By virtue of the exceptionally simple form of { the discrete Wigner function} of $\hat{\rho}_S(\epsilon)$, given in equation \eqref{wigner-function}, the complete weight enumerators of Theorem \ref{theorem-1} become simple weight enumerators.

Applying Theorem \ref{theorem-1}(a) to $\hat{\rho}_S(\epsilon)$ we find the following corollary.
\begin{corollary}The probability for successful projection of $\hat{\rho}(\epsilon)^{\otimes n}$, $\nu$, onto $\mathcal S$, is given by the \textbf{simple weight enumerator} of $\mathcal S^\perp$, 
    \begin{equation}
    \tr \left( \hat{\Pi}_{\mathcal S} \hat{\rho}_S(\epsilon)^n \right) = w(\mathcal S^\perp;~ x(\epsilon),y(\epsilon)), 
\end{equation}
where $x$, and $y$ are given by eqs. \eqref{xeps} and \eqref{yeps}. \label{simple-corollary}
\end{corollary}

We can also simplify part (b) of 
Theorem \ref{theorem-1}. Let us assume we {are} distilling using a stabilizer code that has a complete set of transversal Clifford gates. The output state $\hat{\rho}'=f_{MSD}(\hat{\rho})$ of such a distillation protocol will then also be of the form in equation \eqref{wigner-function}, with parameter $\epsilon'$. (Alternatively, if the stabilizer code does not have a complete set of Clifford gates, one could also twirl the output state by symplectic rotations to bring it into this form.) Define $x'$ and $y'$ via $W(\hat{\rho}';0,0)=x'$ and $W(\hat{\rho}';i,j)=y'$ for $(i,j) \neq (0,0)$. Using Theorem \ref{theorem-1}(b), we see $x'$ and $y'$ are given by,
\begin{eqnarray}
x' & = & \frac{w(\mathcal S; ~x,y)}{w(\mathcal S^\perp;~x,y)} \\
y' & = & \frac{1}{8} \frac{w(\mathcal S^\perp;~x,y)-w(\mathcal S;~x,y)}{w(\mathcal S^\perp;~x,y)}. 
\end{eqnarray}
The formal variables $x$ and $y$ used to define the simple weight enumerator in equation \ref{simple-weight-enumerator} are now reinterpreted as entries of the discrete Wigner function in equation \ref{wigner-function}. If we rewrite this expression in terms of the noise parameter $\epsilon$, we find, 
\begin{equation}
    \epsilon' = 3\frac{3A\left(z(\epsilon)\right)+B\left(z(\epsilon)\right)}{4B\left(z(\epsilon)\right)}. \label{msd-formula}
\end{equation}
where, 
\begin{equation}
    z(\epsilon) = \frac{y(\epsilon)}{x(\epsilon)} = \frac{3-\epsilon}{8 \epsilon-6}, \label{zeps}
\end{equation}
$A(z)=w(\mathcal S; 1,z)$ and $B({z})=w(\mathcal S^\perp; 1,z)$. We have thus characterized the noise reduction of a distillation protocol for the strange state in terms of its simple weight enumerators $A(z)$ and $B(z)$.

\subsubsection*{Example: the $[[11,1,5]]_3$ Golay code}
To illustrate the above formalism, let us apply it to the 11-qutrit Golay code of \cite{2020golay}. The 11-qutrit Golay code is an $[[11,1,5]]_3$ CSS-code formed using two copies of the (self-dual) classical ternary Golay code. Its weight enumerator is computed (e.g., via Magma \cite{magma}) to be:
\begin{equation}
    A(z) = 1+528 z^6+7920 z^8+11000 z^9+23760 z^{10}+15840 z^{11}.
\end{equation}
Using the MacWilliams identity, we find,
\begin{equation}
    B(z) = 1+528 z^5+528 z^6+15840 z^7+40920 z^8+129800 z^9+198000 z^{10}+145824 z^{11}.
\end{equation} 
Substitute these results into equation \eqref{msd-formula}, to obtain
{\small 
\begin{equation}
    \begin{split}
        \epsilon' & =  \frac{3 \left(48336 z^{11}+67320 z^{10}+40700 z^9+16170 z^8+3960 z^7+528 z^6+132 z^5+1\right)}{145824 z^{11}+198000 z^{10}+129800 z^9+40920 z^8+15840 z^7+528 z^6+528 z^5+1} \\
         & =  \left(3021 \epsilon ^{11}-24816 \epsilon ^{10}+92180 \epsilon ^9-203280 \epsilon ^8+292710 \epsilon ^7-283536 \epsilon ^6+181764 \epsilon ^5-71280 \epsilon ^4+13365 \epsilon ^3\right)/ \\
        & \phantom{=} \big(990 \epsilon ^{11}-7920 \epsilon ^{10}+27500 \epsilon ^9-50490 \epsilon ^8+37620 \epsilon ^7+47256 \epsilon ^6-172656 \epsilon ^5+243540 \epsilon ^4-204930 \epsilon ^3 \\ & +106920 \epsilon ^2-32076 \epsilon +4374\big) \\
       & \approx    55 \epsilon^3/18 + O(\epsilon^4). \label{golayeq}
    \end{split}
\end{equation}
}

The threshold of the code, $\epsilon_*$, is the critical value of $\epsilon$ such that, $\epsilon<\epsilon_*$ implies $\epsilon'<\epsilon$. Using \eqref{golayeq}, we find 
\begin{equation}
\begin{split}
    \epsilon_* & =\frac{1}{45} \left(-262 \sqrt[3]{\frac{2}{405 \sqrt{109}-2981}}+2^{2/3} \sqrt[3]{405 \sqrt{109}-2981}+31\right) \\
    & \approx 0.387.
\end{split}
\end{equation}
Interestingly, $z_*=z(\epsilon_*)$ satisfies a simple cubic equation, $11 z_*^3+12 z_*^2+3 z_*+1=0$.

\subsection{Conditions for magic state distillation}

There are two conditions that a stabilizer code must satisfy for it to qualify as a magic state distillation routine for the strange state. First observe that the limit $\epsilon \to 0$ of pure strange states corresponds to $z(\epsilon=0)=-1/2$. We first require that the probability of successful projection to be nonzero in the limit $\epsilon \to 0$. This translates into the requirement
\begin{equation}
    B\left( z( \epsilon=0) \right) =B(-1/2) \neq 0. \label{condition-1}
\end{equation}
We also require that the noise suppression be better than linear.

Assuming equation \eqref{condition-1} is satisfied, the noise-suppression exponent, $\delta$, of the magic state distillation routine, $\epsilon' = \Theta(\epsilon^\delta)$, is determined by the smallest power of $\epsilon$ that divides $3A(z(\epsilon))+B(z(\epsilon))$. Let us write
\begin{equation}
    3A(z(\epsilon)) + B(z(\epsilon)) = C_0 + C_1 \epsilon + C_2 \epsilon^2 + \ldots 
\end{equation} 
Generically, we expect $C_0$ and $C_1$ will be non-zero. The necessary and sufficient conditions for $C_0$ and $C_1$ to vanish are,
\begin{equation}
\begin{split}
    \left( 3A(z(\epsilon)) + B(z(\epsilon)) \right)\bigg|_{\epsilon=0}             & = 0, \\
    \frac{d}{d\epsilon}\bigg|_{\epsilon=0}  \left( 3A(z(\epsilon)) + B(z(\epsilon)) \right) & = 0.
\end{split}
\end{equation}
Translated into $z$, these conditions become
\begin{eqnarray}
    3A(-1/2)+B(-1/2) & = & 0, \label{condition-0} \\
   3A'(-1/2) + B'(-1/2) & = & 0 . \label{condition-2}
\end{eqnarray}
As a check, observe that, for the weight enumerators of the 11-qutrit Golay code, $A(-1/2)=2187/64$, $B(-1/2)=6561/64$, $A'(-1/2)=-8019/16$ and $B'(-1/2)=24057/16$, these conditions are satisfied. 

For $n$ odd, equation \eqref{condition-0} is automatically satisfied, by virtue of the MacWilliams identity, which simplifies at $z=-1/2$ to:
\begin{equation}
    B(-1/2)= 3(-1)^n A(-1/2).
\end{equation} 

We can also ask, when do we get cubic noise suppression? If condition in equation \eqref{condition-2} is satisfied, the condition for $C_2$ to vanish {is} 
\begin{equation}
    3 A''(-1/2)+B''(-1/2)=0. \label{three}
\end{equation} 
For $n$ odd, this condition is automatically {satisfied} whenever equation \eqref{condition-2} holds, by virtue of the MacWilliams identities. Taking derivatives of the MacWilliams identity at $z=-1/2$, we find that,
\begin{eqnarray}
    B'(-1/2) & = & (-1)^{n+1} \left(3 A'(-1/2)+8n A(-1/2) \right), \label{macwilliams1}\\
    B''(-1/2) & = & \frac{1}{3} (-1)^n \left(9 A''(-1/2)+48 (n-1) A'(-1/2)+64(n^2-n)A(-1/2)\right). \label{macwilliams2}
\end{eqnarray}
When $n$ is odd, it is straightforward to see that the MacWilliams identities above along with equation \eqref{condition-2} imply that equation \eqref{three} is satisfied. Thus, cubic noise suppression is guaranteed for any odd-length stabilizer code whose weight enumerator satisfies the two conditions, \eqref{condition-1} and \eqref{condition-2}.

\section{Search for distillation routines}
\label{sec:computational-search}
With the above results in place, a computational search for distillation routines for the qutrit strange state is straightforward. For each stabilizer code $\mathcal S$ in our search space, we compute the simple weight enumerator $A(z)$, and then $B(z)$ using the MacWilliams identity. We then check if conditions \eqref{condition-1} and \eqref{condition-2} are satisfied.

\subsection{Narrowing the search space via symmetry}
\label{sec:symmetry}
When are two stabilizer codes equivalent for the purposes of magic state distillation? Conventionally, one considers two quantum error-correcting codes to be equivalent if they differ by local Clifford operations. However, Theorem \ref{theorem-1} shows that the output of a generic magic state distillation routine depends on the complete weight enumerators of $\mathcal S$ and its cosets, which are not invariant under local Clifford transformations. 
Therefore, two stabilizer codes which differ by local Clifford transformations may, in general, give rise to different magic state distillation protocols. Indeed, if $\mathcal S$ and $\mathcal{\tilde{S}}$ differ by a local Clifford unitary $C$, then distilling with $\mathcal{ \tilde{S}}$ is equivalent to first acting with $C$ then distilling with $\mathcal S$. 
Acting with a local Clifford $C$ prior to distillation will, in general, induce an error on a magic state $\ket{M}$, unless $\ket{M}$ is an eigenvector of $C$.
\footnote{As a very simple example of this, consider the 5-qubit code of \cite{MSD}, $\mathcal S_5$, defined to be generated by $\{ XZZXI, IXZZX, XIXZZ, ZXIXZ \}$, which distills Bravyi and Kitaev's $\ket{T}$ state. If we conjugate the first qubit with the Clifford unitary $YH$, we obtain the stabilizer code $\mathcal {\tilde{S}}_5$ generated by $\{ -ZZZXI, IXZZX, -ZIXZZ, -XXIXZ \}$. $\mathcal {\tilde{S}}_5$ clearly does not distill $\ket{T}$ states. To see this, note that the Clifford $YH$ maps $\ket{T}$ directly opposite to $\ket{T}$ on the Bloch sphere, and therefore acting with $YH$ prior to distillation induces an additional error.}

This observation increases the size of our search space substantially -- to search for all magic state distillation routines associated with a given stabilizer code $\mathcal S$, we need to search over all orbits of $\mathcal S$ under local Clifford transformations, and all possible eigenvalues of the $n-1$ stabilizers. For each $[[n,1]]_3$ code, this could increase the search space by a factor as large as, $|SL(2,\mathbb Z_3)|^n 3^{n-1}=24^n 3^{n-1}$. It is possible, however, to substantially reduce this search space size, by demanding that both the stabilizer codes we study and the magic states we wish to distill possess certain symmetries.

Suppose the magic state we wish to distill is invariant under a subgroup $G$ of the single-qudit Clifford group. It is natural to require\footnote{One, in principle, can also consider magic state distillation without twirling the input states -- see \cite{rall2017fractal}.} our noisy input states first undergo a twirling procedure so that they are also invariant under $G$, as described in detail in \cite{jain2020qutrit}. Let $C$ be a single-qudit Clifford-unitary, if a twirled noisy magic state $\hat{\rho}_M$, satisfies $C\hat{\rho}_M (C)^\dagger=\hat{\rho}_M$, then two stabilizer codes which differ via local Clifford transformations belonging to the subgroup of the Clifford group generated {by} $C$ are equivalent for distillation of the $\hat{\rho}_M$ magic state. 

The group $G$ has been computed for various qutrit and ququint  magic states in \cite{jain2020qutrit}. There, it was shown that the qutrit magic state $\ket{S}$ is a simultaneous eigenvector of all symplectic rotations, as mentioned earlier in section \ref{sec:prelim}. Assuming noisy input magic states are twirled by applying a random symplectic rotation, two stabilizer codes are equivalent for $\ket{S}$-state distillation if they are related to each other by local symplectic transformations. This result is reflected in the fact that Corollary \ref{simple-corollary} depends only on the simple weight enumerator, not the complete weight enumerator.

It is also natural to require that all of the unitaries in $G$ be transversal gates for the stabilizer code used for distillation. Our motivation for this is as follows. Suppose $G$ is generated by the Clifford unitaries $\{ C_i\}$. Suppose our desired magic state $\ket{M}$ is the unique simultaneous eigenvector of all the $\{C_i\}$, which is true for the magic states in \cite{jain2020qutrit}. Then $C_i \ket{M}=\lambda_i \ket{M}$. If, for all $i$, $C_i^{\otimes n}$ acts as the logical operator $\bar{C}_i^{\pm 1}$, then $\bar{C}_i^{\pm 1} \ket{M}^{\otimes n}=\lambda_i^{n}\ket{M}^{\otimes n}$. Then demanding $\lambda_i^{n}=\lambda_i^{\pm 1}$ for all $C_i$, ensures that $\Pi_S \ket{M}^{\otimes n} \sim \bar{\ket{M}}$. This condition, which can also be used to restrict the size $n$ of $\mathcal S$, therefore ensures that, if $\Pi_S \ket{M}^{\otimes n} \neq 0$, pure magic states decode to pure magic states, although it does not guarantee a noise reduction. This is not a necessary condition for distillation\footnote{It is not satisfied by triorthogonal codes (which possess a transversal non-Clifford gate) \cite{Bravyi_2012}.}, {but it is true for virtually all magic state distillation routines studied in the literature that we are aware of, e.g., the 5-qubit code in \cite{MSD} and the 11-qutrit Golay code in \cite{2020golay}. This condition also allows us to place restrictions on the size of the codes expected to distill $\ket{M}$. Let the order of $C_i$ be $m_i$. Then we require $n \equiv \pm 1 \mod m_i$.}   

In this paper, we are interested in distilling the qutrit magic state $\ket{S}$, whose  symmetry group $G$ is the set of all symplectic rotations, which, combined with $X$ and $Z$, generate all single-qudit Clifford unitaries. Transversality of symplectic rotations also ensures that $X$ and $Z$ are transversal. Therefore, a natural family of candidate stabilizer codes for distillation of the strange state are $[[n,1]]_3$ stabilizer codes with a complete set of transversal Clifford gates. Symplectic rotations are generated by the Hadamard operator, $\hat{H}$, which has order $4$, and $\hat{Z}^{-1}\hat{S}$, which has order $p=3$. We therefore expect only such codes for which $n \equiv \pm 1 \mod 12$ to be candidates for distillation.

One can, more generally, impose that only a subgroup of $G$ is transversal, but then $\ket{M}$ will, in general, not be the unique eigenvector of the generators of $G$.\footnote{This weaker condition applies, for example, to some of the codes in \cite{DawkinsHoward}.} We may, therefore, instead, impose the less-restrictive requirement that only $\hat{H}^2$ be a transversal gate to obtain a large family of candidate codes, for a more exhaustive search. Notice that, by equation \eqref{h2-action}, acting with $(\hat H^2)^{\otimes n}$ on a stabilizer projector, with a possibly non-trivial syndrome, corresponds to replacing each generator $\hat{D}(\vec{u}_i,\vec{v}_i)$ of the stabilizer code by $\hat{D}(-\vec{u}_i,-\vec{v}_i)$, leaving the eigenvalues unchanged. Alternatively, its action can be thought of as changing the eigenvalues $\omega^{a_i}$ of each generator to $\omega^{-a_i}$. Demanding that $(H^2)^{\otimes n}$ commute with the codespace is therefore equivalent to demanding that the eigenvalues of all stabilizers in the code must be $\omega^0=+1$. Because $\hat{H}^2$ has order $2$, we expect that $n$ should be odd.

We therefore choose our search space to consist of two families of codes:
\begin{enumerate}
\item $[[n,1]]_3$ stabilizer codes with trivial syndrome, and
\item $[[n,1]]_3$ codes with a complete set of transversal Clifford gates.
\end{enumerate} 
We present the results for a search over each of these two families of codes in the next two subsections.

\subsection{Stabilizer codes with trivial syndrome}

We first turn our attention to distillation with all stabilizer codes with trivial syndrome. We can enumerate all such codes for $n\leq 9$ using the classification of codes in \cite{Danielsen2009, Danielsen2012}. To do this, we make use of the correspondence between stabilizer codes and additive codes over $GF(p^2)$ \cite{calderbank1998quantum, ketkar2006nonbinary}. Recall that the stabilizers of any $[[n,k]]_p$ stabilizer code form an additive, self-orthogonal code over $GF(p^2)$ of the form $(n,p^{n-k})_{GF(p^2)}$.  \cite{Danielsen2009, Danielsen2012} classified $(n,3^{n})$ self-orthogonal additive codes over $GF(9)$, of size $n\leq 10$, using graph-theoretic techniques. These correspond to $[[n,0]]_3$ stabilizer codes. 

We require a classification of all $[[n,1]]_3$ stabilizer codes, which correspond, instead, to additive $(n,3^{n-1})_{GF(9)}$ codes. We can obtain such a classification from the results of \cite{Danielsen2009, Danielsen2012} using  a standard construction in classical coding theory known as \textit{shortening}, described in many textbooks, such as \cite{MacWilliamsSloane}. To shorten a code, we choose a particular coordinate $i \in (1, \ldots , n)$ of the code and remove all codewords that are non-zero on the $i$th coordinate; we then delete the $i$th coordinate to obtain a code of length $n-1$. It is easy to see that this operation preserves additivity and self-orthogonality. Shortening an additive $(n,p^{n-k})_{GF(p^2)}$ code yields an $(n-1, p^{n-k-1})_{GF(p^2)}$ code. Moreover, \textit{any} $(n-1, p^{n-k-1})_{GF(p^2)}$ code can be obtained by shortening some $(n,p^{n-k})_{GF(p^2)}$ code. In the language of stabilizer projectors, shortening a code corresponds to a form of channel-state duality \cite{NielsenChuang2010} -- the projector onto an $[[n,0]]_p$ stabilizer code 
$\mathcal S$ can be written as 
\begin{equation}
\Pi = \sum_{k_1=0}^p \sum_{k_2=0}^p  \cdots \sum_{k_n=0}^p  A_{k_1 k_2 \ldots k_n} \bra{k_1 k_2 \ldots k_n}.
\end{equation} 
Shortening this code at the coordinate $1$, gives rise to an $[[n-1,1]]_p$ code $S'$, whose projector can be written as, 
\begin{equation}
\Pi' = \sum_{k_1=0}^p \sum_{k_2=0}^p  \cdots \sum_{k_n=0}^p  A_{k_1 k_2 \ldots k_n} \ket{\bar{k}_1}\bra{k_2 \ldots k_n},
\end{equation}
where $\ket{\bar{k}}$ corresponds to the logical $\ket{k}$ state in the $[[n-1,1]]_p$ code for some choice of logical operators. The stabilizers of $\mathcal S'$ are precisely those stabilizers of $\mathcal S$ that act trivially on the first qudit.

Therefore, by enumerating all inequivalent ways of shortening the codes classified in  \cite{Danielsen2009, Danielsen2012}, we can construct all $[[n,1]]_3$ stabilizer codes with $n\leq 9$. This is straightforward to do, using, e.g., MAGMA \cite{magma}. The classification of \cite{Danielsen2012} also includes those $(12,3^{12})_{GF(9)}$ codes corresponding to $[[12,0,6]]_3$ stabilizer states, from which we can construct some, but not all, $[[11,1]]_3$ stabilizer codes.

We thus searched over all $[[n,1,d]]_3$ codes for ${n} \leq 9$, and all $[[11,1,d]]_3$ codes that can be obtained from applying the shortening operation to a $[[12,0,6]]_3$ stabilizer state. Remarkably{,} the only code we found that could distill the qutrit strange state with better-than-linear {noise suppression} was the 11-qutrit Golay code. There were also a few $[[9,1]]_3$ and $[11,1]]_3$ codes that distilled the strange state with linear noise suppression, (i.e., $\epsilon' = \alpha \epsilon$ with $\alpha<1$), all of which had lower thresholds than the 11-qutrit Golay code.

\subsection{{Stabilizer codes with a complete set of transversal Clifford gates}}
{We now turn our attention to codes with a complete set of transversal Clifford gates. 
To do this, we use the following lemma, which relates codes with a complete set of transversal Clifford gates to CSS codes. Although this lemma, or at least its analogue for qubits \cite{gottesman1997stabilizer}, may be well-known to some readers, we include a proof for completeness.
\begin{lemma}
    Any $[[n,1]]_p$ stabilizer code with a complete set of transversal Clifford gates must be a CSS code, formed from two copies of a maximal self-orthogonal classical $p$-ary code. \label{CSS-lemma}
\end{lemma} 
\begin{proof}
First recall some basic facts about CSS codes \cite{calderbank1996good, steane1996multiple}. A CSS code is a stabilizer code of the form 
\begin{equation}
    \mathcal S_{\rm CSS}=\mathcal S_X \oplus \mathcal S_Z,
\end{equation} where each generator of $M^{(X)}_i$ of $\mathcal S_X$ is of the form $M^{(X)}_i=\hat{D}(\vec{u}_i,0)$ and each generator $M^{(Z)}_i$ of $\mathcal S_Z$ is of the form $M^{(Z)}_i=\hat{D}(0,\vec{v}_i)$. Equivalently, a code $\mathcal S$ is a CSS-code if, for any $M=\hat{D}(\vec{u},\vec{v}) \in \mathcal S$, $D(\vec{u},0) \in \mathcal S$ and $D(0,\vec{v}) \in \mathcal S$. $\mathcal S_X$ and $\mathcal S_Z$ can each be thought of as classical codes over $\mathbb Z_p$, with $\mathcal S_X \subset \mathcal S_Z^\perp$ and $\mathcal S_Z \subset \mathcal S_X^\perp$. The lengths of $\mathcal S_X$ and $\mathcal S_Z$ are both equal to $n$, and $\dim \mathcal S_X + \dim \mathcal S_Z=n-k$. If $\mathcal S_Z= \mathcal S_X$ in a CSS-code, then $\mathcal S_Z$ must be self-orthogonal. If, in addition, $\mathcal S_{CSS}$ is an $[[n,1]]_p$ code, then $2 \dim \mathcal S_Z=n-1$. Then $n$ must be odd, and $\mathcal S_Z$ is an $[n,\lfloor \frac{n}{2} \rfloor]_3$ self-orthogonal code. It is a well-known fact from classical coding theory that any self-orthogonal code with these parameters is maximal, see, e.g., \cite{TernarySelfOrthogonal1, MacWilliamsSloane}.

Let $\mathcal S$ be a stabilizer code with a complete set of transversal Clifford gates. If $M=\hat{D}(\vec{u},\vec{v}) \in \mathcal S$, then, by transversality, so is $\hat{M}'=(V_F^{-1})^{\otimes n} M V_F^{\otimes n} =\hat{D}(\vec{u}',\vec{v}')$ where, 
\begin{equation}
\begin{pmatrix} u_i' \\ v_i' \end{pmatrix} = F \begin{pmatrix} u_i \\ v_i \end{pmatrix},
\end{equation} for some $F \in SL(2,\mathbb Z_p)$. Choosing $F$ appropriately, we see that $M'=\hat{D}(\vec{u}+\vec{v},-\vec{u})\in \mathcal S$, and $M''=\hat{D}(-\vec{v} , \vec{u})\in \mathcal S$. Then $M'''=M'M'' = \hat{D}(\vec{u},0) \in S$. Similarly, $M^{(4)}=\hat{D}(0,\vec{u}) \in S$. The code is therefore CSS, with $\mathcal S_Z=\mathcal S_X$.
\end{proof}

By Lemma \ref{CSS-lemma}, a search over $[[n,1]]_3$ stabilizer codes with a complete set of transversal Clifford gates therefore translates into a search over CSS codes generated from maximal self-orthogonal ternary codes of odd $n$. Explicitly, if $G_c$ is the generator matrix of the self-orthogonal classical ternary code with $[n, \lfloor \frac{n}{2} \rfloor]_3$, the quantum CSS code is given by the  symplectic matrix:
\begin{equation}
    H = \left( \begin{array}{c|c}
    G_c & 0 \\
   0 & G_c  
    \end{array}\right). 
\end{equation}}
Classical maximal self-orthogonal ternary codes up to size $n=23$ have been classified in \cite{TernarySelfOrthogonal1, TernarySelfOrthogonal2, TernarySelfOrthogonal3, TernarySelfOrthogonal4, TernarySelfOrthogonal5, TernarySelfOrthogonal6}, and are conveniently available on a website maintained by Harada and Munemasa \cite{munemasa_codes_website}. We computed the weight enumerators of all CSS-codes constructed this way from the codes given in \cite{munemasa_codes_website}, using Magma \cite{magma}. We found that no indecomposable\footnote{A code is said to be indecomposable if its generator matrix cannot be written as the direct sum of two smaller generator matrices.} 13, 15, 17 or 19-qutrit CSS codes were able to distill the strange state. 

Somewhat surprisingly, however, we found a total of 646 inequivalent indecomposable 23-qutrit CSS codes\footnote{The 646 inequivalent CSS codes gave rise to 263 different simple weight enumerators.} that were able to distill the strange state with cubic noise suppression. A complete list of the $646$ classical ternary codes that gave rise to these codes is included in MAGMA format \cite{magma} as ancillary data along with the arXiv submission of this paper \cite{prakash2024search}.

There are a total of 1928 indecomposable maximally-self-orthogonal $[23,11]_3$ codes, so the probability that a randomly chosen code will give rise to a quantum CSS code that distills the strange state is $0.335$, which is very close to $1/3$.  The probability that a randomly chosen maximally self-orthogonal $[11,5]_3$ code gives rise to a CSS code that distills the strange state is also $1/3$. This seems to suggest that quantum {error-correcting} codes that distill the qutrit strange state are actually quite common.

\begin{figure}
    \centering
    \includegraphics[width=0.6 \textwidth]{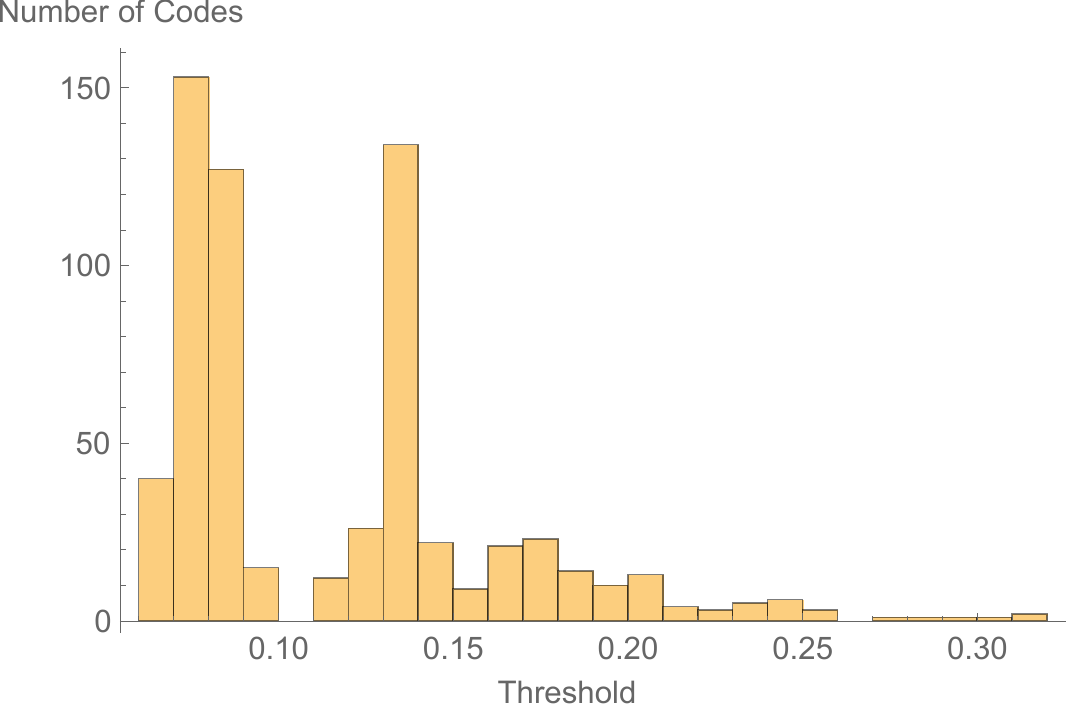}
    \caption{A histogram of all the thresholds that arise from 23-qubit CSS codes that are able to distill the strange state. We found a total of 646 codes, and the highest threshold was $\epsilon_* = 0.318$.}
    \label{fig:thresholds}
\end{figure}

\begin{figure}[h]
    \centering
    \includegraphics[width=0.6\textwidth]{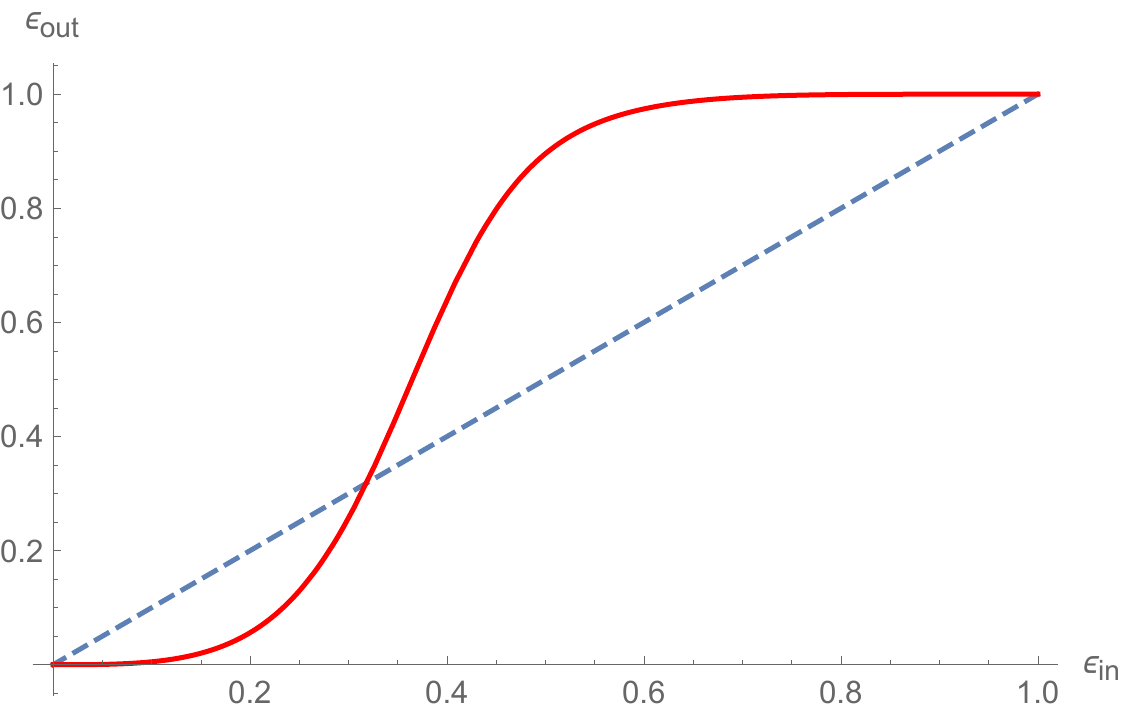}
    \caption{The  distillation performance of the $[[23,1,5]]_3$ code defined via equation \eqref{23-code}.}
    \label{plot23}
\end{figure}

The thresholds that arise from these codes range from $0.063$ to $0.318$, and are plotted in a histogram in Figure \ref{fig:thresholds}. None of these thresholds exceed that of the ternary Golay code. {The $23$-qutrit code with the highest threshold was a $[[23,1,5]]_3$ code formed from two copies of the classical $[23, 11, 6]_3$ code with generator matrix:}
\begin{equation}
    G_C = \begin{pmatrix}
     1 & 0 & 0 & 0 & 0 & 0 & 0 & 0 & 0 & 0 & 0 & 1 & 2 & 2 & 0 & 0 & 2 & 0 & 2 & 0 & 0 & 0 & 0\\
 0 & 1 & 0 & 0 & 0 & 0 & 0 & 0 & 0 & 0 & 0 & 2 & 2 & 1 & 1 & 2 & 1 & 2 & 1 & 2 & 2 & 0 & 1\\
 0 & 0 & 1 & 0 & 0 & 0 & 0 & 0 & 0 & 0 & 0 & 1 & 0 & 0 & 0 & 0 & 1 & 1 & 0 & 0 & 2 & 2 & 0\\
 0 & 0 & 0 & 1 & 0 & 0 & 0 & 0 & 0 & 0 & 0 & 1 & 0 & 0 & 0 & 0 & 1 & 2 & 0 & 1 & 0 & 1 & 0\\
 0 & 0 & 0 & 0 & 1 & 0 & 0 & 0 & 0 & 0 & 0 & 0 & 0 & 0 & 2 & 0 & 0 & 1 & 0 & 2 & 2 & 2 & 0\\
 0 & 0 & 0 & 0 & 0 & 1 & 0 & 0 & 0 & 0 & 0 & 2 & 2 & 0 & 0 & 2 & 1 & 0 & 2 & 0 & 0 & 0 & 0\\
 0 & 0 & 0 & 0 & 0 & 0 & 1 & 0 & 0 & 0 & 0 & 1 & 0 & 0 & 2 & 0 & 1 & 0 & 0 & 2 & 0 & 2 & 0\\
 0 & 0 & 0 & 0 & 0 & 0 & 0 & 1 & 0 & 0 & 0 & 1 & 0 & 0 & 1 & 0 & 1 & 0 & 0 & 0 & 1 & 1 & 0\\
 0 & 0 & 0 & 0 & 0 & 0 & 0 & 0 & 1 & 0 & 0 & 0 & 0 & 0 & 2 & 0 & 0 & 1 & 0 & 1 & 1 & 0 & 1\\
 0 & 0 & 0 & 0 & 0 & 0 & 0 & 0 & 0 & 1 & 0 & 0 & 0 & 2 & 1 & 2 & 0 & 2 & 1 & 2 & 2 & 0 & 1\\
 0 & 0 & 0 & 0 & 0 & 0 & 0 & 0 & 0 & 0 & 1 & 1 & 2 & 2 & 1 & 1 & 2 & 2 & 1 & 2 & 2 & 0 & 1 
 \end{pmatrix}. \label{23-code}
\end{equation}
This quantum CSS code has weight enumerator, 
\begin{equation}
\begin{split} 
A(z) & =  2079023616 z^{23}+6035662080 z^{22}+8258226816 z^{21}+7208904960 z^{20}+4504066560 z^{19}+2182781824 z^{18} \\ &+ 790797312 z^{17}+252077184 z^{16}+52015680 z^{15}+14590080 z^{14}+2083104 z^{13}+628800 z^{12}+121824 z^{11} \\ & +58320 z^{10}+16120 z^9+4608 z^8+720 z^6+1,
\end{split}
\end{equation}
which gives rise to a distillation performance 
\begin{equation}
    \epsilon' \approx \frac{73 \epsilon ^3}{18}+O\left(\epsilon ^4\right),
\end{equation}
plotted in Figure \ref{plot23}. 

\begin{figure}[h]
    \centering
    \includegraphics[width=0.8\textwidth]{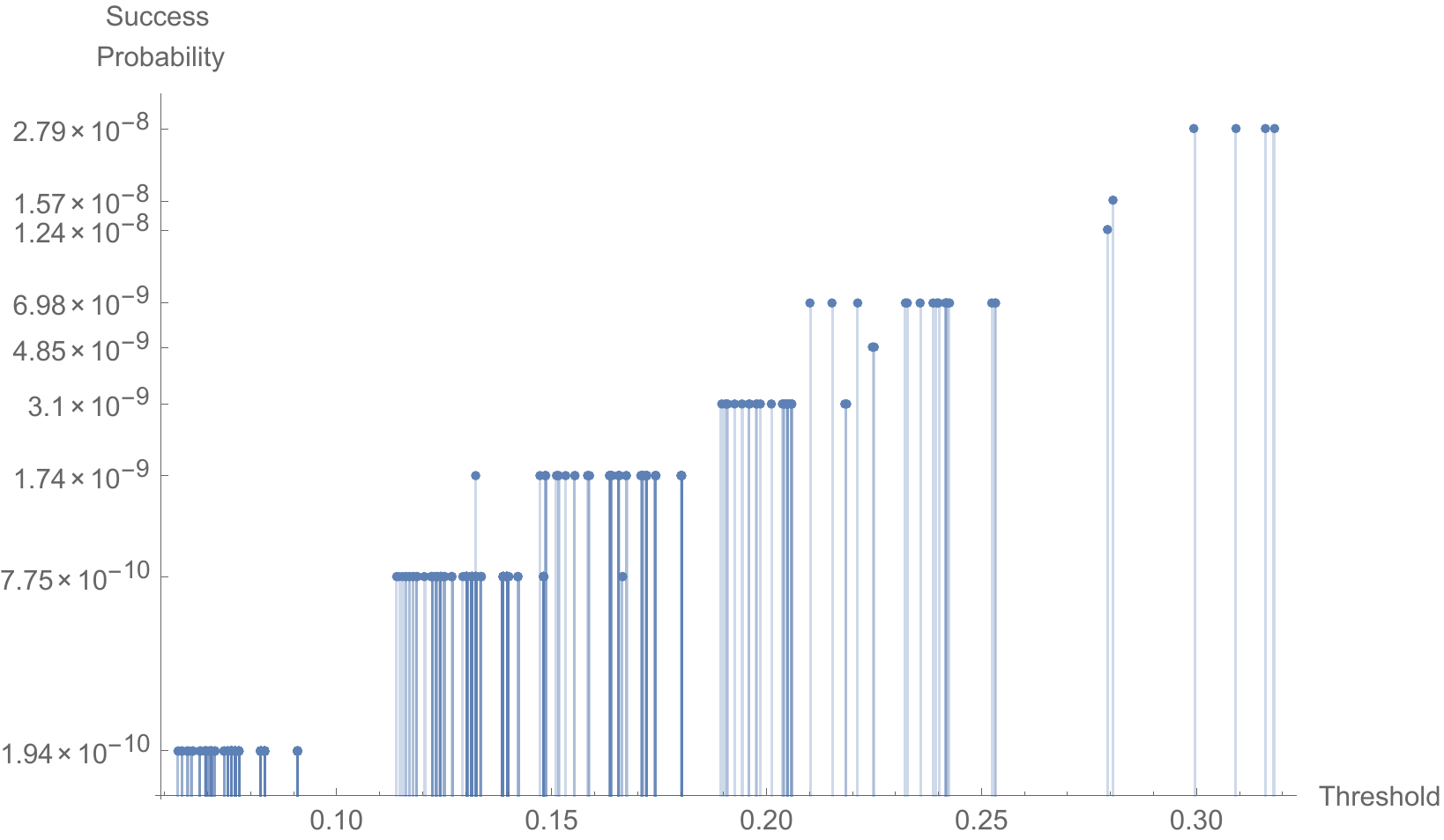}
    \caption{A scatter plot of the success probability (logarithmic scale) and threshold for the {646} CSS codes that distill the strange state.}
    \label{success-threshold}
\end{figure}

The probability of successful projection onto the codespace for these codes is very low. The success probabilities range from $1/5159780352 \approx 1.9 \times 10^{-10}$ to ${1}/{35831808} \approx 2.8 \times 10^{-8}$.  The highest probability of successful distillation is attained for the 23-qutrit code in equation \eqref{23-code}, and is ${1}/{35831808}$ -- three other codes have the same success probability and very similar thresholds. It appears that the success probability is correlated with the threshold. A plot of success probability versus threshold for the 646 codes that distill the strange state is shown in Figure \ref{success-threshold}. 

\section{Discussion}
\label{sec:discussion}

This paper is motivated by the conjecture \cite{nature} that contextuality is sufficient for universal quantum computation. This translates directly into the conjecture that any state outside the Wigner polytope can be distilled into a pure magic state. A systematic search for magic state distillation routines for the qutrit strange state $\ket{S}$, which lies directly above the centre of one of the faces of the qutrit Wigner polytope, is a direct probe of this fundamental conjecture. In this work, we developed a simple formalism for studying magic state distillation routines for the qutrit strange state via weight enumerators. This formalism enabled us to carry out an extensive search for distillation routines, covering $n$-to-$1$ distillation routines, for $n$ as large as $23$. 

In the introduction, we posed some open questions regarding distillation of the qutrit strange state, which our computational search enables us to answer. 
\begin{itemize}
\item \textit{Do stabilizer codes, other than the 11-qutrit ternary Golay code, distill the strange state?} Yes. We have demonstrated the existence of over 600 codes that distill the strange state with cubic noise suppression. 

\item \textit{How do their thresholds compare?} While the existence of codes that distill the strange state appears generic, the high threshold of the 11-qutrit Golay code is not. None of the codes we found that distill the strange state possess a threshold exceeding that of the 11-qutrit Golay code. 
\end{itemize}
Do our results support the existence of a sequence of distillation routines with threshold approaching the limit set by contextuality? Our finding that $1/3$ of all $[[11,1]]_3$ and $[[23,1]]_3$ stabilizer codes with a complete set of transversal Clifford gates distill the strange state with cubic noise suppression, provides some non-trivial evidence for the possibility of such a sequence. However, we have no concrete evidence for a sequence of codes with \textit{increasing} thresholds; and finding examples of codes that distill the strange state with thresholds exceeding that of the 11-qutrit Golay code seems extremely challenging. 

This work leads to several new questions. Is there a code that distills the strange state with better threshold, or with better-than-cubic noise suppression? Can one understand the observation that exactly $1/3$ of $[[23,1]]_3$ distill the strange state with cubic noise suppression? The study of weight enumerators of classical self dual and maximal-self orthogonal codes is a rich subject -- can ideas from invariant theory, applied to weight enumerators, as in \cite{kalra2025invarianttheorymagicstate}, be used to say anything about the conjecture that contextuality is sufficient for universal quantum computation? 

Magic state distillation \cite{MSD, Knill2004QuantumCW} is a somewhat mysterious application of quantum error-correcting codes. For the $\ket{T}$ state of \cite{MSD} few distillation routines are known, and the mechanism behind distillation remains unclear. Distillation of the qutrit strange state seems as mysterious as that of the $\ket{T}$-state, though it is perhaps even less well-understood. However, the results of this paper suggest that
the study of distillation of the strange state may in fact be more tractable than that of the qubit $\ket{T}$ state, thanks to the simple relation between distillation performance and simple enumerators valid only for the qutrit strange state. We hope that, by enlarging the landscape of codes known to distill this state, {this} will lead, in the future, to a better understanding of magic state distillation.

\section*{Acknowledgements}

SP thanks Prof. P.S. Satsangi for inspiration and guidance. We also thank Mark Howard and Amolak Kalra for comments on a draft of this manuscript. We acknowledge the support of MeitY QCAL, Amazon Braket and the Anusandhan National Research Foundation (formerly DST-SERB) grant CRG/2021/009137.

\newpage

\appendix

\section*{Appendices}
\section{{Stabilizer projectors in the discrete phase space formalism}} \label{appendix-A}
In this appendix, we prove some technical lemmas {that} describe the action of stabilizer projectors on phase-point operators used to define discrete phase space. These lemmas are needed for the proof of Theorem \ref{theorem-1} in the main text. {These lemmas may be self-evident to readers well-versed in the discrete phase space formalism \cite{Wootters1987, Gibbons:2004dij, Gross} but we include explicit proofs for completeness.} 

\begin{lemma}
{Let $\mathcal S$ be a stabilizer code consisting of a group of phase-free commuting $n$-qudit Heisenberg-Weyl operators, and let 
$$\hat{\Pi}_{\mathcal S}=\frac{1}{p^{n-k}} \sum_{M \in \mathcal S} M$$ be the projector onto its codespace with trivial syndrome.} Then,
\begin{eqnarray}
   \tr \left(  \hat{\Pi}_{\mathcal S} \hat{A}(\vec{u},\vec{v}) \right)
   & = & \begin{cases} 0 & (\vec{u}|\vec{v}) \notin \mathcal S^\perp \\
   1 & (\vec{u}|\vec{v}) \in \mathcal S^\perp
   \end{cases}.
\end{eqnarray} \label{phase-point-lemma}
\end{lemma}

\begin{proof}

Let $\{ M_1,~M_2, \ldots M_{n-k} \}$ be any choice of $n-k$ independent generators of $\mathcal S$. The stabilizer group $\mathcal S$ splits the $p^n$ Hilbert space into $p^{(n-k)}$ different $p^k$-dimensional subspaces. Each subspace may be labeled by a vector of syndromes $\vec{s}=(s_1, \ldots, s_{n-k})$, and denotes the subspace that satisfies
\begin{equation}
    M_i \ket{\psi} = \omega^{s_i} \ket{\psi}, \quad \forall i \in (1, \ldots, n-k). 
\end{equation}
The vector $\vec{s}=(0, \ldots, 0)=0$ denotes the trivial subspace. Let $\hat{\Pi}^{\vec{s}}_{\mathcal S}$ denote the projector onto the subspace labeled by $\vec{s}$. Explicitly, we can write 
\begin{equation}
    \hat{\Pi}^{\vec{s}}_{\mathcal S} = {\frac{1}{p^{n-k}} \prod_{i=1}^{n-k} \sum_{j=0}^{p-1}\left(\omega^{-s_i}M_i\right)^j =}\frac{1}{p^{n-k}} \sum_{M \in \mathcal S} \omega^{f(\vec{s},M)} M,
\end{equation}
for some phases $\omega^{f(\vec{s},M)}$ that depend on the syndrome $\vec{s}$, with $f(\vec{s},M)\in \mathbb Z_p$. 

Observe the following properties of $f(\vec{s},M)$. By definition, $f(0,M)=0$. For $\vec{s} \neq 0$, $f(\vec{s},M)$ takes on each of the values $j \in \mathbb Z_p$, i.e.,  $\{ 0,~1, ~\ldots,~p-1\}$ an equal number of times, i.e.,
\begin{equation}
    |\{ M~| f(\vec{s},M)=j \}|=p^{n-k-1}.
\end{equation}
Note also that, if $M$ is a phase-free Heisenberg-Weyl operator, then 
\begin{equation}
\tr \left(\hat{\Pi}^{\vec s}_{\mathcal S} M\right) = \begin{cases}
    0 & M^{-1} \notin \mathcal S \\
    p^k \omega^{f(\vec{s},M^{-1})} & M^{-1} \in \mathcal S
\end{cases}.
\end{equation}
Therefore 
\begin{eqnarray}
    \tr \left(\hat{\Pi}^{\vec{s}}_{\mathcal S} \hat{A}(0,0)^{\otimes n} \right)  & = & \frac{1}{p^n}  \sum_{\vec{u},\vec{v}} \tr \left(\hat{\Pi}^{\vec{s}}_{\mathcal S} \hat{D}(\vec{u},\vec{v}) \right) \\
    & = & \frac{1}{p^n}  \sum_{M \in \mathcal S} \omega^{f(\vec{s},M^{-1})}{p^{k}}  \\
    & = & \begin{cases} 
    \frac{p^k}{p^n}\cdot p^{n-k-1} \left( 1 + \omega + \omega^2 + \ldots \omega^{p-1} \right) =0 & \vec{s}\neq 0 \\
    1 & \vec{s}=0
    \end{cases} \label{A00-lemma}
\end{eqnarray}

Note that 
\begin{equation}
    \hat{D}(\vec{u},\vec{v})^{-1}\hat{\Pi}^0_{\mathcal S} \hat{D}(\vec{u},\vec{v}) = \hat{\Pi}^{\vec{s}}_{\mathcal S}, \label{projector-to-syndrome}
\end{equation}
for some syndrome $\vec{s}$, which is trivial (i.e. $\vec{s}=0$) if $\hat{D}(\vec{u},\vec{v})$ commutes with all $M \in \mathcal S$, and non-trivial (i.e., $\vec{s} \neq 0$) otherwise. The condition that $\hat{D}(\vec{u},\vec{v})$ commutes with all $M \in \mathcal S$ is equivalent to the condition $(\vec{u},\vec{v}) \in \mathcal S^\perp$. Therefore,
\begin{eqnarray}
   \tr \hat{\Pi}_{\mathcal S} \hat{A}(\vec{u},\vec{v})
   & = & \tr \left( \hat{D}(-\vec{u},-\vec{v}) \hat{\Pi}_S \hat{D}(\vec{u},\vec{v}) \hat{A}(0,0)^{\otimes n} \right) \\
   & = & \begin{cases} \tr (\hat{\Pi}_S^{\vec{s}\neq 0} \hat{A}(0,0)^{\otimes n}) & (\vec{u}|\vec{v}) \notin \mathcal S^\perp \\
   \tr (\hat{\Pi}^0_S \hat{A}(0,0)^{\otimes n}) & (\vec{u}|\vec{v}) \in \mathcal S^\perp
   \end{cases},
\end{eqnarray}
and the result follows using equation \eqref{A00-lemma}.
\end{proof}

A corollary to the above lemma that we will also use is 
\begin{lemma} \label{dual-lemma}
    Let $\mathcal S^\perp$ be the dual of $\mathcal S$. Then, if we define
\begin{equation}
    \hat{\Pi}_{\mathcal S^\perp} \equiv \frac{1}{p^{2k}}\sum_{\vec x, \vec z \in \mathbb Z_p^k} \bar{D}(\vec{x},\vec{z}) \hat{\Pi}_{\mathcal S} =  \frac{1}{p^{n+k}} \sum_{(\vec{u}|\vec{v}) \in \mathcal S^\perp} \hat{D}(\vec{u},\vec{v}),  \label{pseudo-projector}
\end{equation}
we have,
\begin{eqnarray}
   \tr \left(  \hat{\Pi}^0_{\mathcal S^\perp} \hat{A}(\vec{u},\vec{v}) \right)
   & = & \begin{cases} 0 & (\vec{u}|\vec{v}) \notin \mathcal S \\
   1 & (\vec{u}|\vec{v}) \in \mathcal S
   \end{cases}.
\end{eqnarray}
\end{lemma}

Note that, for an $[[n,k]]_p$ stabilizer code with $k \geq 1$, $\mathcal S \subset \mathcal S^\perp$. While $\mathcal S$ is self-orthogonal, $\mathcal S^\perp$ is not.  This means that all the $\hat{D}(u,v)$ in the sum on the RHS of equation \eqref{pseudo-projector} do not commute, and the operator $\hat{\Pi}_{\mathcal S^\perp}$ is \textit{not} a projector. Therefore the proof of Lemma \ref{phase-point-lemma} does not immediately apply to this case.
    
\begin{proof}
    For simplicity, assume that we are working with an $[[n,1]]_p$ stabilizer code. Choose a logical operator $\bar{D}(u,v)=D(\vec{u}_L,\vec{v}_L)$.  Let $\mathcal S \cup [\bar{D}(u,v)]$ denote the group of phase-free Heisenberg-Weyl operators generated by $\mathcal S$ and the logical operator $\bar{D}(u,v)=D(\vec{u}_L,\vec{v}_L)$. Clearly, 
    \begin{equation}
        \Pi_{\mathcal S + (\vec{u}_L,\vec{v}_L)}=\frac{1}{p}\left(1+\bar{D}(u,v)+\bar{D}(u,v)^2 + \ldots \right) \hat{\Pi}^0_{\mathcal S}. 
    \end{equation}
    Note that $\mathcal S \cup [\bar{D}(u,v)]$ is a stabilizer code that satisfies the conditions of Lemma \ref{phase-point-lemma}, so we can apply it to obtain, 
    \begin{eqnarray}
   \tr \left(  \hat{\Pi}^0_{\mathcal S + (\vec{u}_L,\vec{v}_L)} \hat{A}(\vec{u},\vec{v}) \right)
   & = & \begin{cases} 0 & (\vec{u}|\vec{v}) \notin \mathcal S + (\vec{u}_L,\vec{v}_L)\\
   1 & (\vec{u}|\vec{v}) \in \mathcal S + (\vec{u}_L,\vec{v}_L)
   \end{cases}. \label{logical-op}
   \end{eqnarray}

   There are $p^2-1$ distinct logical operators $\bar{D}(u,v)$ not equal to the logical identity operator. These can be divided into $(p+1)$ families of $(p-1)$ commuting operators \cite{Wootters1987,Gross,appleby2008spectra}. (For example, one such family is, $\{ \bar{D}(0,1),~\bar D(0,2),~ \ldots \bar D(0,p-1) \}$.) Index these families using $L=1, \ldots, p+1$, and choose one operator from each family. Then, adding equation \eqref{logical-op} for each of these operators, we have, 
   \begin{equation}
    \sum_{L=1}^{p+1} \tr \left(  \hat{\Pi}^0_{\mathcal S + (\vec{u}_L,\vec{v}_L)} \hat{A}(\vec{u},\vec{v}) \right)   =  \begin{cases} 0 & (\vec{u}|\vec{v}) \notin \mathcal S^\perp \\
   1 & (\vec{u}|\vec{v}) \in \mathcal S^\perp - \mathcal S \\
   p+1 & (\vec{u}|\vec{v}) \in \mathcal S
   \end{cases}. \label{big-trace}
   \end{equation}

   Note that
   \begin{equation}
      \sum_{L=1}^{p+1}  \hat{\Pi}^0_{\mathcal S + (\vec{u}_L,\vec{v}_L)} =  p\hat{\Pi}^0_{\mathcal S^\perp}+ \hat{\Pi}^0_{\mathcal S},
   \end{equation}
because $\bar{D}(0,0)$ appears $p+1$ times in the sum, on the LHS of \eqref{big-trace}. Therefore \begin{equation}
    \tr \left( \hat{\Pi}^0_{\mathcal S^\perp} \hat{A}(\vec u,\vec v) \right)=  \frac{1}{p}  \sum_{L=1}^{p+1} \tr \left( \hat{\Pi}^0_{\mathcal S + (\vec{u}_L,\vec{v}_L)} \hat{A}(\vec u,\vec v) \right)-  \tr \left( \hat{\Pi}^0_{\mathcal S}\hat{A}(\vec u,\vec v) \right)
\end{equation}
and the corollary then follows using \eqref{big-trace} and Lemma \ref{phase-point-lemma}.
\end{proof}

Though we will not use this result in the paper, it is straightforward to modify Lemma \ref{phase-point-lemma} to handle the case of non-trivial syndromes as well. 

\begin{lemma} \label{non-trival-syndromes}
Let $\mathcal S$ be a stabilizer code consisting of a group of phase-free commuting $n$-qudit Heisenberg-Weyl operators, and let 
$\hat{\Pi}^{\vec{s}}_{\mathcal S}$ be the projector onto its codespace with non-trivial syndrome. Then, there exists $(\vec{u}_s|\vec{v}_s) \notin \mathcal S^\perp$, such that
\begin{eqnarray}
   \tr \left( \hat{\Pi}^{\vec{s}}_{\mathcal S} \hat{A}(\vec{u},\vec{v}) \right)
   & = & \begin{cases} 0 & (\vec{u}|\vec{v}) + (\vec{u}_s|\vec{v}_s) \notin \mathcal S^\perp \\
   1 & (\vec{u}|\vec{v}) + (\vec{u}_s|\vec{v}_s) \in \mathcal S^\perp
   \end{cases}.
\end{eqnarray} \label{phase-point-lemma-v2}
\end{lemma}
\begin{proof}
For any syndrome $\vec{s}$ there exists some Pauli-operator $D(\vec{u}_s,\vec{v}_s) \notin \mathcal S^\perp$ such that,
\begin{equation}
    \hat{\Pi}^{\vec{s}}_{\mathcal S}=\hat{D}(\vec{u}_s,\vec{v}_s)\hat{\Pi}^0_{\mathcal S} \hat{D}(\vec{u}_s,\vec{v}_s)^{-1}. 
\end{equation}
Therefore, 
\begin{eqnarray}
   \tr \hat{\Pi}^{\vec{s}}_{\mathcal S} \hat{A}(\vec{u},\vec{v})
   & = & \tr \left( \hat{D}(-\vec{u}_s,-\vec{v}_s) \hat{\Pi}^0_S \hat{D}(\vec{u}_s,\vec{v}_s) \hat{D}(\vec{u},\vec{v})  \hat{A}(0,0)^{\otimes n} \hat{D}(-\vec{u},-\vec{v}) \right) \\
   &= &  \tr \hat{\Pi}^{0}_{\mathcal S} \hat{A}(\vec{u}+\vec{u}_s,\vec{v}+\vec{v}_s)\\
   & = & \begin{cases} 0 & (\vec{u}|\vec{v})+(\vec{u}_s|\vec{v}_s) \notin \mathcal S^\perp \\
   1 & (\vec{u}|\vec{v})+(\vec{u}_s|\vec{v}_s) \in \mathcal S^\perp.
   \end{cases}
\end{eqnarray}
\end{proof}

Lemma \ref{phase-point-lemma-v2} gives rise to a modified, slightly messier, version of Theorem \ref{theorem-1} with each vector $(\vec{u}, \vec{v})$ shifted by $(\vec{u}_s| \vec{v}_s)$. Interestingly, the discrete phase space formalism allows one to avoid the use of signed weight enumerators \cite{rall2017signed} altogether when working with qudits of odd-prime dimension.

\section{Some codes that do not distill the strange state}
\label{Appendix-B}
In \cite{SharmaGarani2024}, it was recently proposed that two qutrit stabilizer codes -- a $[[13,1]]_3$ and a $[[29,1]]_3$ CSS code -- distill the strange state. \cite{SharmaGarani2024} claimed their thresholds to be $0.425$ and $\approx 0.7$, respectively. Using the techniques developed in section \ref{sec:weight-enumerators}, we were able to compute the distillation performance of these codes exactly, by directly computing their simple weight enumerators in Magma \cite{magma}. We found that neither of these two codes distill the strange state.\footnote{An erratum to \cite{SharmaGarani2024} has subsequently been issued \cite{erratum}, however, we include this appendix for reference.} 

\subsection{The $[[13,1,4]]_3$ code}

\begin{figure}
    \centering
    \includegraphics[width=0.8 \textwidth]{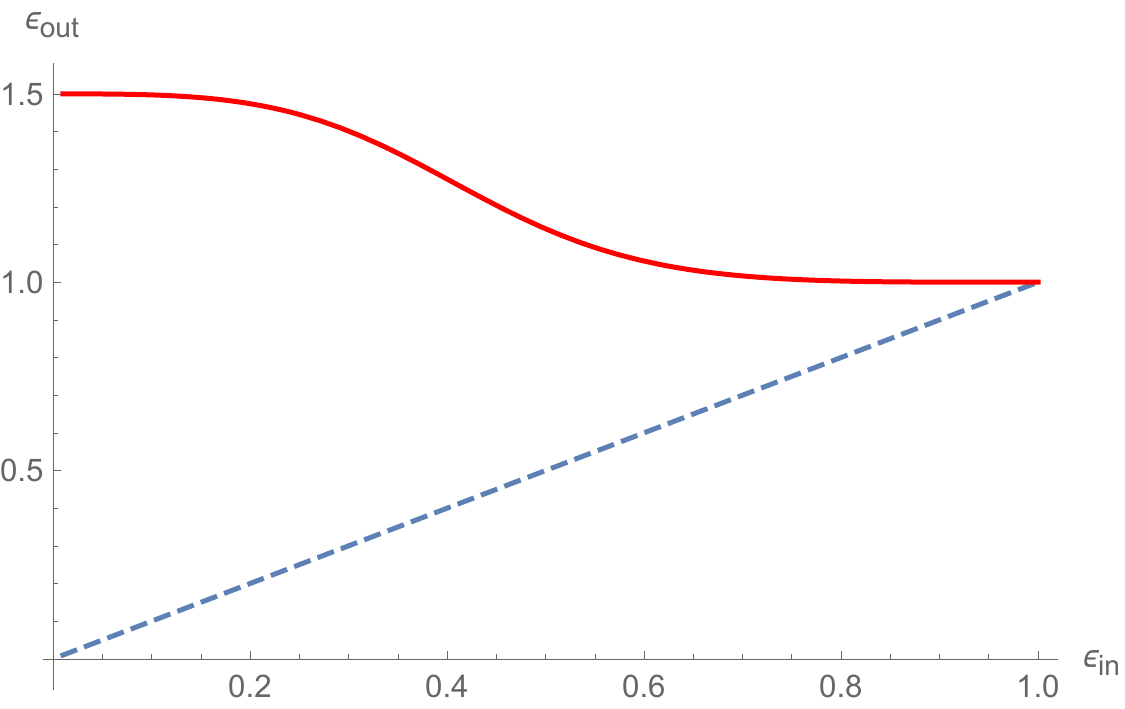}
    \caption{The distillation performance for the $[[13,1,4]]_3$ code of \cite{SharmaGarani2024}. For small $\epsilon_{\rm in}$, we find $\epsilon_{\rm out} \to 3/2$, which, from Equation \eqref{mixed}, corresponds to a mixture of two states orthogonal to the strange state.}
    \label{plot13}
\end{figure}

The $[[13,1 ,4]]_3$ CSS-code of \cite{SharmaGarani2024} generated from two copies of the $[13,6,3]_3$ maximal self-orthogonal ternary code with generator matrix is, in row-reduced form:
\begin{equation}
M_{13}= \begin{pmatrix}
1 &	0 &	0 &	0 &	0 &	0 &	0 &	0 &	2 &	0 &	2 &	0 &	0\\
0 &	1 &	0 &	0 &	0 &	0 &	0 &	2 &	2 &	1 &	1 &	2 &	0\\
0 &	0 &	1 &	0 &	0 &	0 &	1 &	2 &	0 &	1 &	0 &	2 &	2\\
0 &	0 &	0 &	1 &	0 &	0 &	1 &	0 &	2 &	0 &	1 &	2 &	2\\
0 &	0 &	0 &	0 &	1 &	0 &	0 &	0 &	2 &	2 &	1 &	1 &	1\\
0 &	0 &	0 &	0 &	0 &	1 &	2 &	1 &	2 &	0 &	1 &	1 &	0
\end{pmatrix}.
\end{equation}
The weight enumerator of the $[[13,1 ,4]]_3$ code is
\begin{equation}
    A(z) = 1+8 z^3+600 z^6+720 z^7+4320 z^8+18320 z^9+61200
   z^{10}+151200 z^{11}+178144 z^{12}+116928 z^{13}.
\end{equation}
This gives rise to a relation 
\begin{equation}
    {\epsilon' = \frac{3}{2}-\frac{20 \epsilon ^3}{9}+O\left(\epsilon ^4\right),}
\end{equation}
plotted in Figure \ref{plot13}. As $\epsilon \to 0$, we see that $\epsilon' \to 3/2$, which corresponds to the mixed state $\frac{1}{2}\ket{\psi}\bra{\psi}+\frac{1}{2}\ket{0}\bra{0}$, where $\ket{\psi}=\frac{1}{\sqrt{2}}(\ket{1}+\ket{2})$. This shows that $\ket{S}^{\otimes 13}$ is orthogonal to the codespace of the stabilizer code, so the code is completely unsuitable for magic state distillation. This can {also} be seen from the fact that $B(-1/2)=0$. This feature was shared by $6$ out of the $7$ possible $13$-qutrit codes that arise from our construction using the classical ternary codes listed in \cite{munemasa_codes_website}.

\subsection{The $[[29,1,7]]_3$ code}

The $[[29, 1, 7]]_3$ CSS code of \cite{SharmaGarani2024} is constructed from two copies of a $[29,15]_3$ classical ternary self-orthogonal code, whose parity-check matrix is presented in Figure 8 of \cite{SharmaGarani2024}; for brevity, we do not reproduce it here. Computing its weight enumerator using Magma \cite{magma} (which took approximately one day of computation time on the desktop computer we had access to), we find that the $[[29,1,7]]_3$ CSS code has simple weight enumerator
\begin{equation}
\begin{split}
A(z) = & 1+40 z^6+4280 z^9+96 z^{10}+2832 z^{11}+196584
   z^{12}+198768 z^{13}+1773408 z^{14}+15542368
   z^{15} \\ & +91797024 z^{16}  +565547232 z^{17}+3037545272
   z^{18}+13979050848 z^{19}+55970778960
   z^{20}+192507694176 z^{21} \\ & +559711606992
   z^{22}+1361197350960 z^{23}+2723501140720
   z^{24}+4358977591776 z^{25} \\ & +5363568387600
   z^{26}+4767481212256 z^{27}+2724627154368
   z^{28}+751557878400 z^{29}.
\end{split}
\end{equation} 

\begin{figure}
    \centering
    \includegraphics[width=0.8 \textwidth]{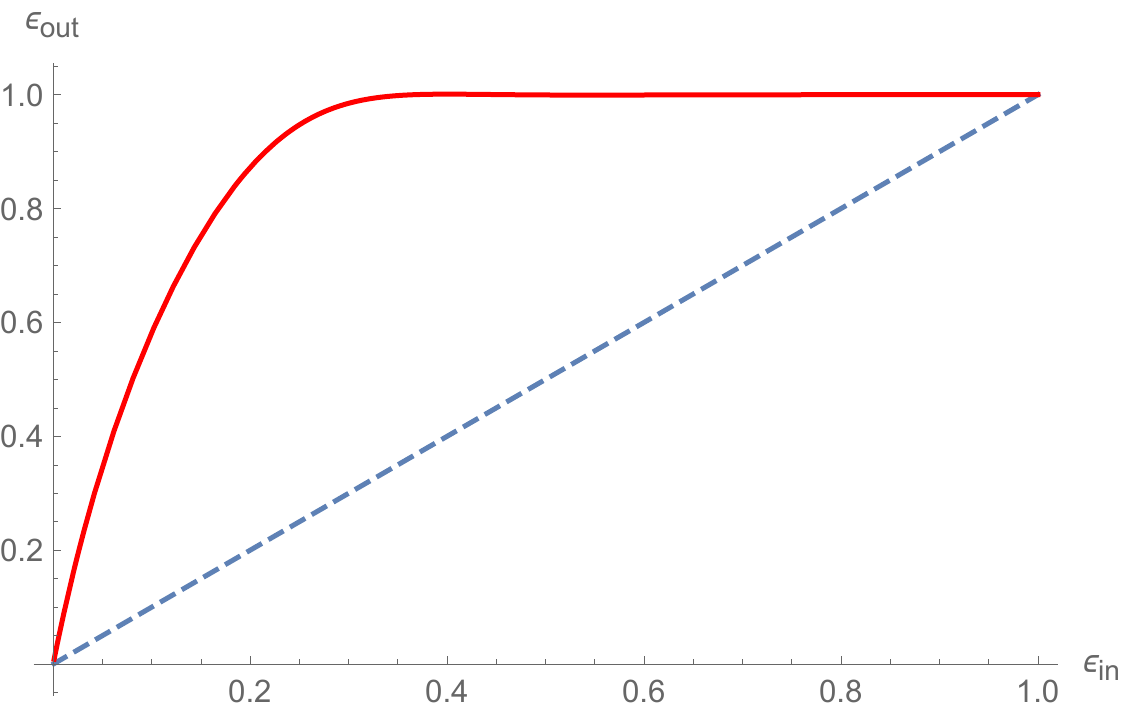}
    \caption{The distillation performance for the $[[29,1,7]]_3$ code of \cite{SharmaGarani2024}. The threshold for distillation is zero.}
    \label{plot29}
\end{figure}

Using this weight-enumerator, we find that the noise reduction of the magic state distillation routine is
\begin{equation}
    \epsilon' = \frac{1937 \epsilon}{224}+O\left({\epsilon}^2\right). \label{distillation29}
\end{equation} 
Equation \eqref{distillation29} is plotted in Figure \ref{plot29}. It is clear from both the Figure and equation \eqref{distillation29} that the threshold for distillation is zero.
Therefore, at present, the $11$-qutrit Golay code has the highest threshold for distillation of the strange state. It would be interesting to look at other 29-qutrit CSS codes, however, to our knowledge, a complete list of $[29, 14]_3$  self-orthogonal ternary codes is not yet available.

\bibliographystyle{ssg}
\bibliography{qudit2024}

\end{document}